\documentclass[12pt,preprint]{aastex}  
\usepackage{graphicx}
\usepackage{txfonts}
\newcommand{\va}{v_{\mathrm{A 0}}}
\newcommand{\cs}{c_{\mathrm{s 0}}}

\newcommand{\nct}{\tilde{c}_{\mathrm{T 0}}}

\newcommand{\lambdamod}{\tilde{\Lambda}_0}

\newcommand{\pd}{\partial}
\newcommand{\td}{\tau_{\mathrm{D}}}

\newcommand{\Omegao}{\Omega_0}

\begin{document}

	\title{Propagation of nonadiabatic magnetoacoustic waves in a threaded prominence with mass flows}

	\shorttitle{MHD waves in prominence multi-thread systems}

   \author{R. Soler, R. Oliver, and J. L. Ballester}

   \affil{Departament de F\'isica, Universitat de les Illes Balears,
              E-07122, Palma de Mallorca, Spain}
              \email{[roberto.soler;ramon.oliver;joseluis.ballester]@uib.es}

  \begin{abstract}
High resolution observations of solar filaments suggest 
the presence of groups of prominence threads, i.e. the fine-structures of 
prominences, which oscillate coherently (in phase). In addition, mass flows 
along threads have been often observed. Here, we investigate the effect of 
mass flows on the collective fast and slow nonadiabatic magnetoacoustic wave 
modes supported by systems of prominence threads. Prominence fine-structures are modeled as parallel, homogeneous and infinite cylinders embedded in a coronal environment. The magnetic field is uniform and parallel to the axis of threads. Configurations of identical and nonidentical threads are both explored.  We apply the $T$-matrix theory of acoustic scattering to obtain the oscillatory frequency and the eigenfunctions of linear magnetosonic disturbances. We find that the existence of wave modes with a collective dynamics, i.e. those that produce significant perturbations in all threads, is only possible when the Doppler-shifted individual frequencies of threads are very similar. This can be only achieved for very particular values of the plasma physical conditions and flow velocities within threads. 
  \end{abstract}


%________________________________________________________________

\section{Introduction}

Prominences/filaments are fascinating coronal magnetic structures, whose dynamics and properties are not well-understood yet. The long life of the so-called quiescent prominences (several weeks) suggests that the cool and dense prominence material is maintained against gravity and thermally shielded from the much hotter and much rarer solar corona by means of some not well-known processes. However, it is believed that the magnetic field must play a crucial role in both the support and isolation of prominences. High-resolution observations of solar filaments reveal that they are formed by a myriad of horizontal structures called threads \citep[e.g.,][]{lin2005}, which have been observed in the spines and barbs of both active region and quiescent filaments \citep{lin2008}. The width of these fine-structures is typically in the range 0.2~arcsec --  0.6~arcsec, which is close to the resolution of present-day telescopes, whereas their lengths are between 5~arcsec and 20~arcsec \citep{lin2004}. Threads are assumed to be the basic substructures of filaments and to be aligned along magnetic field lines. From the point of view of theoretical modeling, prominence threads are interpreted as large coronal magnetic flux tubes, with the denser and cooler (prominence) region located at magnetic field dips that correspond to the observed threads. Although some theoretical works have attempted to model such structures \citep[e.g.,][]{ballesterpriest, schmitt, rempel, heinzel06}, there are some concerns about their formation and stability that have not been resolved yet.

Small amplitude oscillations, propagating waves and mass flows are some phenomena usually observed in prominences and prominence threads \citep[see some recent reviews by][]{oliverballester02, ballester, banerjee}. Periods of small-amplitude prominence oscillations cover a wide range from less than a minute to several hours, and they are usually attenuated in a few periods \citep{molowny, terradasobs}. Focusing on prominence threads, some works have detected oscillations and waves in such fine-structures \citep[e.g.,][]{yi1, yi2, lin2004, okamoto,lin2007}. In particular, \citet{yi1} and \citet{lin2007} suggested the presence of groups of near threads that moved in phase, which may be a signature of collective oscillations. On the other hand, mass flows along magnetic field lines have been also detected \citep{zirker94, zirker, lin2003, lin2005}, with typical flow velocities of less than 30~km~s$^{-1}$ in quiescent prominences, although larger values have been detected in active region prominences \citep{okamoto}. Regarding the presence of flows, a phenomenon which deserves special attention is the existence of the so-called counter-streaming flows, i.e., opposite flows within adjacent threads \citep{zirker, lin2003}. 

Motivated by the observational evidence, some authors have broached the theoretical investigation of prominence thread oscillations by means of the magnetohydrodynamic (MHD) theory in the $\beta = 0$ approximation. First, some works \citep{joarder,diaz2001,diaz2003} focused on the study of the ideal MHD oscillatory modes supported by individual nonuniform threads in Cartesian geometry. Later, \citet{diaz2002} considered a more representative cylindrical thread and obtained more realistic results with respect to the spatial structure of perturbations and the behavior of trapped modes. Subsequently, the attention of authors turned to the study of collective oscillations of groups of threads, and the Cartesian geometry was adopted again for simplicity. Hence, \citet{diaz2005} investigated the collective fast modes of systems of nonidentical threads and found that the only nonleaky mode corresponds to that in which all threads oscillate in spatial phase. Later, \citet{diazroberts} considered the limit of a periodic array of threads and obtained a similar conclusion. Therefore, these results seem to indicate that all threads within the prominence should oscillate coherently, even if they have different physical properties. However, one must bear in mind that the Cartesian geometry provides quite an unrealistic confinement of perturbations, and so systems of more realistic cylindrical threads might not show such a clear collective behavior. 

The next obvious step is therefore the investigation of oscillatory modes of systems of cylindrical threads. The first approach to a similar problem was done by \cite{luna1}, who considered a system of two identical, homogeneous cylinders embedded in an unlimited corona. Although \cite{luna1} applied their results to coronal loops, they are also applicable to prominence threads. These authors numerically found that the system supports four trapped kink-like collective modes. These results have been analytically re-obtained by \citet{tom}, by considering the thin tube approximation and bicylindrical coordinates. Subsequently, \citet{luna2} made use of the $T$-matrix theory of acoustic scattering to study the collective oscillations of arbitrary systems of non-identical cylinders. Although the scattering theory has been previously applied in the solar context \citep[e.g.,][]{bogdan87,keppens}, the first application to the study of normal modes of magnetic coronal structures has been performed by \citet{luna2}. They concluded that, contrary to the Cartesian case of \citet{diaz2005}, the collective behavior of the oscillations diminishes when cylinders with nonidentical densities are considered, the oscillatory modes behaving in practice like individual modes if cylinders with mildly different densities are assumed.

The present study is based on \citet{luna2} and applies their technique to the investigation of MHD waves in systems of cylindrical prominence threads. Moreover, we extend their model by considering some effects neglected by them. Here, the more general $\beta \neq 0$ case is considered, allowing us to describe both slow and fast magnetoacoustic modes. In addition, the adiabatic assumption is removed and, following previous papers \citep{soler1,solerapj}, the effect of radiative losses, thermal conduction and plasma heating is taken into account. The detection of mass flows in prominences has motivated us to include this effect in our study, and so the presence of flows along magnetic field lines is also considered here. Therefore, the present work extends our recent investigation \citep[][hereafter Paper~I]{solerapj}, which was focused on individual thread oscillations, to the study of collective MHD modes in prominence multi-thread configurations with mass flows. On the other hand, the longitudinal structure of threads \citep[e.g.,][]{diaz2002} is neglected in the present investigation. For this reason, the effect of including a longitudinal variation of the plasma physical conditions within threads should be investigated in a future work. Finally, the prominence multi-thread model developed here could be an useful tool for future seismological applications \citep[similar to that of][]{hinode}.

 This paper is organized as follows. The description of the model configuration and the mathematical method are given in \S~\ref{sec:math}. Then, the results are presented in \S~\ref{sec:results}. First, the case of two identical prominence threads is investigated in \S~\ref{sec:2treads}. Later, this study is extended to a configuration of two different threads in \S~\ref{sec:ntreads}. Finally, our conclusion is given in \S~\ref{sec:conclusions}.

\section{Mathematical method}
\label{sec:math}

Our equilibrium system is made of an arbitrary configuration of $N$ homogeneous and unlimited parallel cylinders, representing prominence threads, embedded in an also homogeneous and unbounded coronal medium. Each thread has its own radius, $a_j$, temperature, $T_j$, and density, $\rho_j$, where the subscript $j$ refers to a particular thread. On the other hand, the coronal temperature and density are $T_c$ and $\rho_c$, respectively. Cylinders are orientated along the $z$-direction, the $xy$-plane being perpendicular to their axis. The magnetic field is uniform and also orientated along the $z$-direction,  ${\mathit {\bf B}}_j=B_j \hat{\bf e}_z$ being the magnetic field in the $j$-th thread, and ${\mathit {\bf B}}_c=B_c \hat{\bf e}_z$ in the coronal medium. In addition, steady mass flows are assumed along magnetic field lines, with flow velocities and directions that can be different within threads and in the corona. Thus, ${\mathit {\bf U}}_j=U_j \hat{\bf e}_z$ represents the mass flow in the $j$-th thread, whereas ${\mathit {\bf U}}_{\rm c}=U_{\rm c} \hat{\bf e}_z$ corresponds to the coronal flow. For simplicity, in all the following expressions a subscript 0 indicates local equilibrium values, while subscripts $j$ or $c$ denote quantities explicitly computed in the $j$-th thread or in the corona, respectively. 

Such as shown in Paper~I, linear nonadiabatic magnetoacoustic perturbations are governed by the next equation for the divergence of the velocity perturbation, $\Delta =  \nabla \cdot {\mathit {\bf v}}_1$,
\begin{equation}
\Upsilon^2_0 \left[ \Upsilon^2_0 - \left( \tilde{\Lambda}_0^2 + \va^2 \right) \nabla^2 \right] \Delta + \tilde{\Lambda}_0^2 \va^2 \frac{\pd^2}{\pd z^2} \nabla^2 \Delta = 0, \label{eq:basic}
\end{equation}
where $\Upsilon_0$ is the following operator,
\begin{equation}
 \Upsilon_0 = \frac{\pd}{\pd t} + U_0 \frac{\pd}{\pd z},
\end{equation}
while  $\va^2 = \frac{B_0^2}{\mu \rho_0}$ is the Alfv\'en speed squared and $\lambdamod^2$ is the nonadiabatic sound speed squared,
\begin{equation}
 \tilde{\Lambda}^2_0 \equiv \frac{\cs^2}{\gamma} \left[ \frac{\left( \gamma-1 \right) \left( \frac{T_0}{p_0} \kappa_{\parallel 0} k_z^2 + \omega_{T 0} - \omega_{\rho 0} \right) + i \gamma \Omega_0}
{\left( \gamma -1 \right) \left( \frac{T_0}{p_0} \kappa_{\parallel 0} k_z^2 + \omega_{T 0} \right) + i \Omega_0} \right], \label{eq:lambda}
\end{equation}
$\cs^2 = \frac{\gamma p_0}{\rho_0}$ and $\gamma$ being the adiabatic sound speed squared and the adiabatic ratio, respectively. Terms with $\kappa_{\parallel 0}$, $\omega_{\rho 0}$, and $\omega_{T 0}$ are related to nonadiabatic mechanisms, i.e. radiative losses, thermal conduction, and heating (see Paper~I for details). Finally, $\Omegao$ is the Doppler-shifted frequency \citep{terra},
\begin{equation}
 \Omegao = \omega - k_z U_0,
\end{equation}
where $\omega$ is the oscillatory frequency and $k_z$ is the longitudinal wavenumber. Considering cylindrical coordinates, namely $r$, $\varphi$, and $z$ for the radial, azimuthal, and longitudinal coordinates, respectively, we can write $\Delta$ in the following form,
\begin{equation}
 \Delta = \psi \left( r, \varphi \right) \exp \left( i \omega t - i k_z z \right), \label{eq:div}
\end{equation}
where the function $ \psi \left( r, \varphi \right)$ contains the full radial and azimuthal dependence. By inserting this last expression into equation~(\ref{eq:basic}), the following Helmholtz equation is obtained,
\begin{equation}
 \nabla_{r \varphi}^2  \psi \left( r, \varphi \right) + m_0^2\,  \psi \left( r, \varphi \right) = 0, \label{eq:hem}
\end{equation}
where $ \nabla_{r \varphi}^2$ is the Laplacian operator for the $r$ and $\varphi$ coordinates, and
\begin{equation}
 m_0^2 = \frac{\left( \Omega_0^2 - k_z^2 \va^2 \right) \left( \Omega_0^2 - k_z^2 \tilde{\Lambda}^2_0 \right)}
{\left( \va^2 + \tilde{\Lambda}^2_0 \right) \left( \Omega_0^2 - k_z^2 \nct^2 \right)}, \label{eq:m0}
\end{equation}
\begin{equation}
\nct^2 \equiv \frac{\va^2 \tilde{\Lambda}^2_0}{\va^2 + \tilde{\Lambda}_0^2}, \label{eq:ct}
\end{equation}
are the radial wave number and the nonadiabatic tube speed squared, respectively. Moreover, due to the presence of nonideal terms $m_0^2$ is a complex quantity. Since nonadiabatic mechanisms produce a small correction to the adiabatic wave modes, $| \Re(m_0^2) | > | \Im(m_0^2) |$ and the dominant wave character depends on the sign of $\Re(m_0^2)$. Here, we investigate nonleaky modes, which are given by $\Re(m_c^2) < 0$. We impose no restriction on the wave character within threads. 

 In order to solve equation~(\ref{eq:hem}), we consider the technique developed by \citet{luna2} based on the study of wave modes of an arbitrary configuration of cylinders by means of the $T$-matrix theory of acoustic scattering. The novelty with respect to the work of \citet{luna2} is that the method is applied here to solve a Helmholtz equation for the divergence of the velocity perturbation (our eq.~[\ref{eq:basic}]) whereas \citet{luna2} considered an equation for the total pressure perturbation in the $\beta = 0$ approximation (their eq.~[1]). The present approach allows us to study the more general $\beta \neq 0$ case, therefore slow modes are also described. In addition, nonadiabatic effects and mass flows are easily included in our formalism. However, the rest of the technique is absolutely equivalent to that of  \citet{luna2}, and therefore the reader is refered to their work for an extensive explanation of the mathematical technique \citep[see also an equivalent formalism in][]{bogdancatta}. We next give a brief summary of the method.

The main difference between our application and that of \citet{luna2} is in the definition of the $T$-matrix elements. These elements are obtained by imposing appropriate boundary conditions at the edge of threads, i.e., at $|{\bf r} - {\bf r}_j | = a_j$, where ${\bf r}_j$ is the radial vector corresponding to the position of the $j$-th thread center with respect to the origin of coordinates. In our case, these boundary conditions are the continuity of the total pressure perturbation, $p_{\rm T}$, and the Lagrangian radial displacement, $\xi_r =- i v_r / \Omegao$. Expressions for these quantities as functions of $\Delta$ and its derivative are,
\begin{equation}
 p_{\rm T} =  i \rho_0 \frac{\left( \Omegao^2 - k_z^2 \lambdamod^2 \right) \left( \Omegao^2 - k_z^2 \va^2 \right)}{\Omegao^3 m_0^2} \Delta,
\end{equation}
\begin{equation}
 \xi_r = i \frac{\left( \Omegao^2 - k_z^2 \lambdamod^2 \right)}{\Omegao^3 m_0^2} \frac{\partial \Delta}{\partial r}.
\end{equation}
Expressions for the rest of perturbations are given in Appendix~A of Paper~I. Thus, in our case the $T$-matrix elements have the following form,
\begin{equation}
 T_{mm}^j = \frac{m_c \rho_j \left( \Omega_j^2 - k_z^2 v_{{\rm A}_j}^2 \right) J_m \left( m_j a_j \right) J_m' \left( m_c a_j \right) - m_j \rho_c \left( \Omega_c^2 - k_z^2 v_{{\rm A}_c}^2 \right) J_m \left( m_c a_j \right) J_m' \left( m_j a_j \right)}
{m_c \rho_j \left( \Omega_j^2 - k_z^2 v_{{\rm A}_j}^2 \right) J_m \left( m_j a_j \right) {H'}_m^{(1)} \left( m_c a_j \right) - m_j \rho_c \left( \Omega_c^2 - k_z^2 v_{{\rm A}_c}^2 \right) H_m^{(1)} \left( m_c a_j \right) J_m' \left( m_j a_j \right)},
\end{equation}
where $H_m^{(1)}$ and $J_m$ are the Hankel function of the first kind and the Bessel function of order $m$, respectively, while the prime denotes the derivative taken with respect to $r$. Note that the denominator of $T_{mm}^j$ vanishes at the normal mode frequencies of an individual thread. This can be easily checked by comparing it to the dispersion relation of a single thread, see equation~(19) of Paper~I, in which Bessel $K_m$ functions are used instead of Hankel functions. The equivalence between both kinds of functions is given in \citet{abram}. Thus, following  \citet{luna2}, the internal $\psi \left( r, \varphi \right)$ field of the $j$-th thread is,
\begin{equation}
 \psi^j_{\rm int} (r,\varphi)= \sum_{m = -\infty}^{\infty} A_m^j J_m \left( m_j  |{\bf r} - {\bf r}_j | \right)e^{i m \varphi_j}, \label{eq:int}
\end{equation}
whereas the external net field is,
\begin{equation}
 \psi_{\rm ext}(r,\varphi) =  - \sum_j \sum_{m=-\infty}^{\infty} 2 \alpha_{2,m}^j T_{mm}^j H_m^{(1)} \left( m_c  |{\bf r} - {\bf r}_j | \right) e^{i m \varphi_j}, \label{eq:ext}
\end{equation}
where $\varphi_j$ is the azimuthal angle corresponding to the position of the $j$-th thread center with respect to the origin of coordinates, and $\alpha_{2,m}^j$ and $A_m^j$ are constants. Equations~(\ref{eq:int}) and (\ref{eq:ext}) allow us to construct the spatial distribution of $\Delta$. Subsequently, the rest of perturbations can be obtained. Finally, the constants $\alpha_{2,m}^j$ form a homogeneous system of linear algebraic equations,
\begin{equation}
\alpha_{2,m}^j +  \sum_{k \neq j} \sum_{n=-\infty}^{\infty} T_{nn}^k  \alpha_{2,n}^k H_{n-m}^{(1)} \left( m_c |{\bf r}_j - {\bf r}_k | \right) e^{i \left( n -m \right) \varphi_{j k}} = 0, \label{eq:system}
\end{equation}
for $-\infty < m < \infty$. Once both integers $m$ and $n$ are truncated to a finite number of terms, the non-trivial (i.e., non-zero) solution of system~(\ref{eq:system}) gives us a dispersion relation for the oscillatory frequency, $\omega$, which is enclosed in the definitions of $m_j$ and $m_c$.

In the next Sections, we apply the method to obtain the oscillatory frequency and the spatial distribution of perturbations of the wave modes supported by prominence thread configurations. We assume that $k_z$ is real, so a complex frequency is obtained, namely $\omega = \omega_{\rm R} + i \omega_{\rm I}$. The imaginary part of the frequency appears due to the presence of nonadiabatic mechanisms. The oscillatory period, $P$, and the damping time, $\td$, are related to the frequency as follows,
\begin{equation}
P = \frac{2 \pi}{| \omega_{\rm R} |}, \qquad \tau_{\rm D} = \frac{1}{\omega_{\rm I}}.
\end{equation}

\section{Results}
\label{sec:results}

\subsection{Configuration of two identical threads}
\label{sec:2treads}

\begin{figure}[!hp]
\centering
\epsscale{0.49}
\plotone{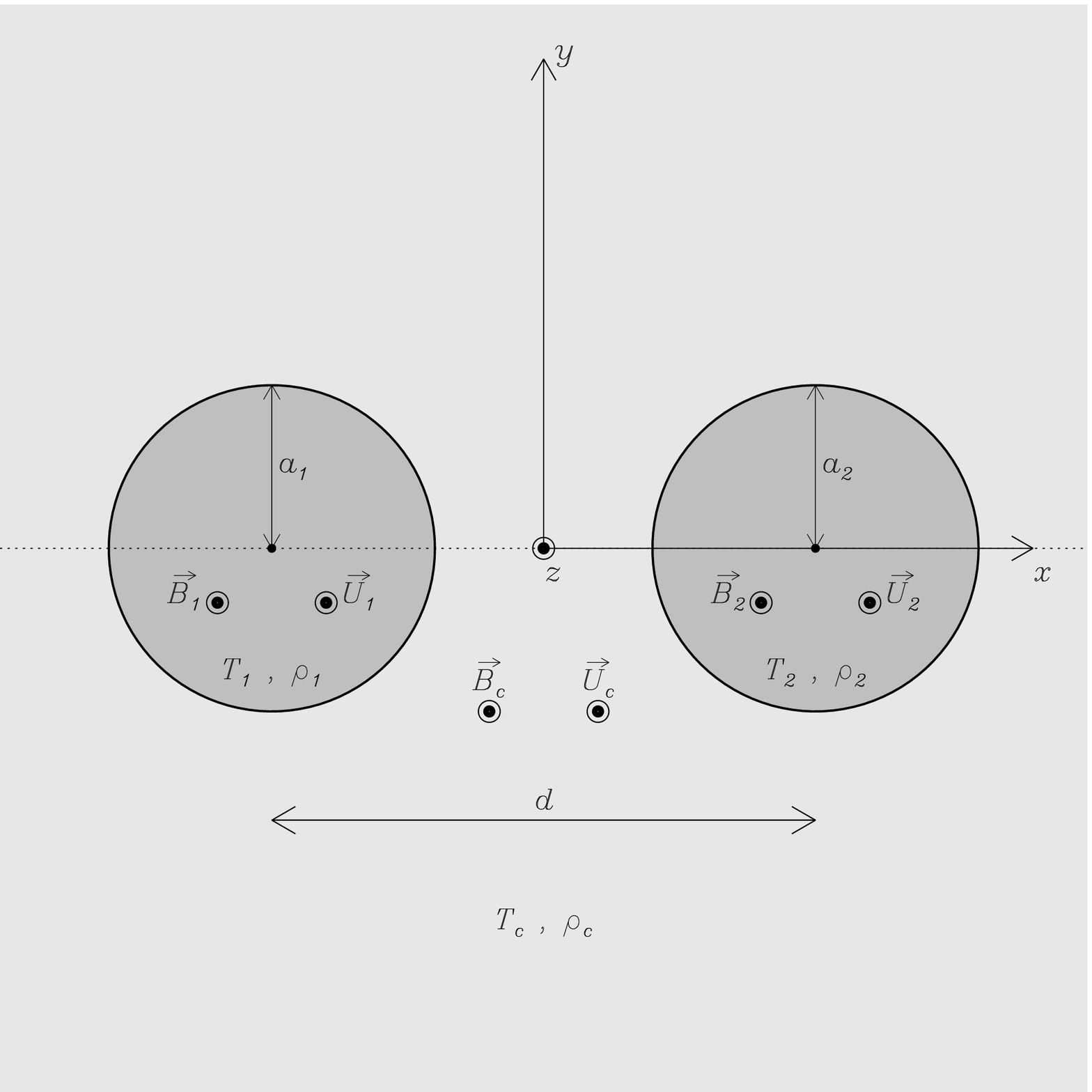}
\caption{Scheme in the $xy$-plane of the model considered in \S~\ref{sec:2treads}. The $z$-axis is perpendicular to the plane of the figure and points towards the reader. \label{fig:model}}
\end{figure}

Now, we consider a configuration of two identical threads (see Fig.~\ref{fig:model}). Their physical conditions are typical of prominences ($T_1 = T_2 =8000$~K, $\rho_1 = \rho_2 = 5 \times 10^{-11}$~kg~m$^{-3}$) while the coronal temperature and density are $T_{\rm c} = 10^6$~K and $\rho_{\rm c} = 2.5 \times 10^{-13}$~kg~m$^{-3}$, respectively. Their radii are $a_1 = a_2 = a =30$~km, and the distance between centers is  $d= 4 a = 120$~km. The magnetic field strength is 5~G everywhere. The flow velocity inside the cylinders is denoted by $U_{\rm 1}$ and  $U_{\rm 2}$, respectively, whereas the flow velocity in the coronal medium is $U_{\rm c}$. Unless otherwise stated, these physical conditions are used in all calculations.

\subsubsection{Wave modes in the absence of flow}

First, we consider no flow in the equilibrium, i.e. $U_{\rm 1} = U_{\rm 2} = U_{\rm c} = 0$. We fix the longitudinal wavenumber to $k_z a = 10^{-2}$, which corresponds to a wavelength within the typically observed range. In addition to the four kink modes described by \citet{luna1}, i.e. the $S_x$, $A_x$, $S_y$, and $A_y$ modes, where $S$ or $A$ denote symmetry or antisymmetry of the total pressure perturbation with respect to the $yz$-plane, and the subscripts refer to the main direction of polarization of motions, we also find two more fundamental collective wave modes (one symmetric and one antisymmetric) mainly polarized along the $z$-direction, which we call $S_z$ and $A_z$ modes following the notation of \citet{luna1}. These new solutions correspond to slow modes which are absent in the investigation of \citet{luna1} due to their $\beta = 0$ approximation. As was stated by \citet{luna2}, a collective wave mode is the result of a coupling between individual modes. So the reader must be aware that in the present work we indistinctly use both expressions, i.e., collective modes and coupled modes, to refer to wave solutions whose perturbations have significant amplitudes in both threads.

\begin{figure}[!hp]
\centering
%\epsscale{0.325}
\epsscale{0.45}
\plotone{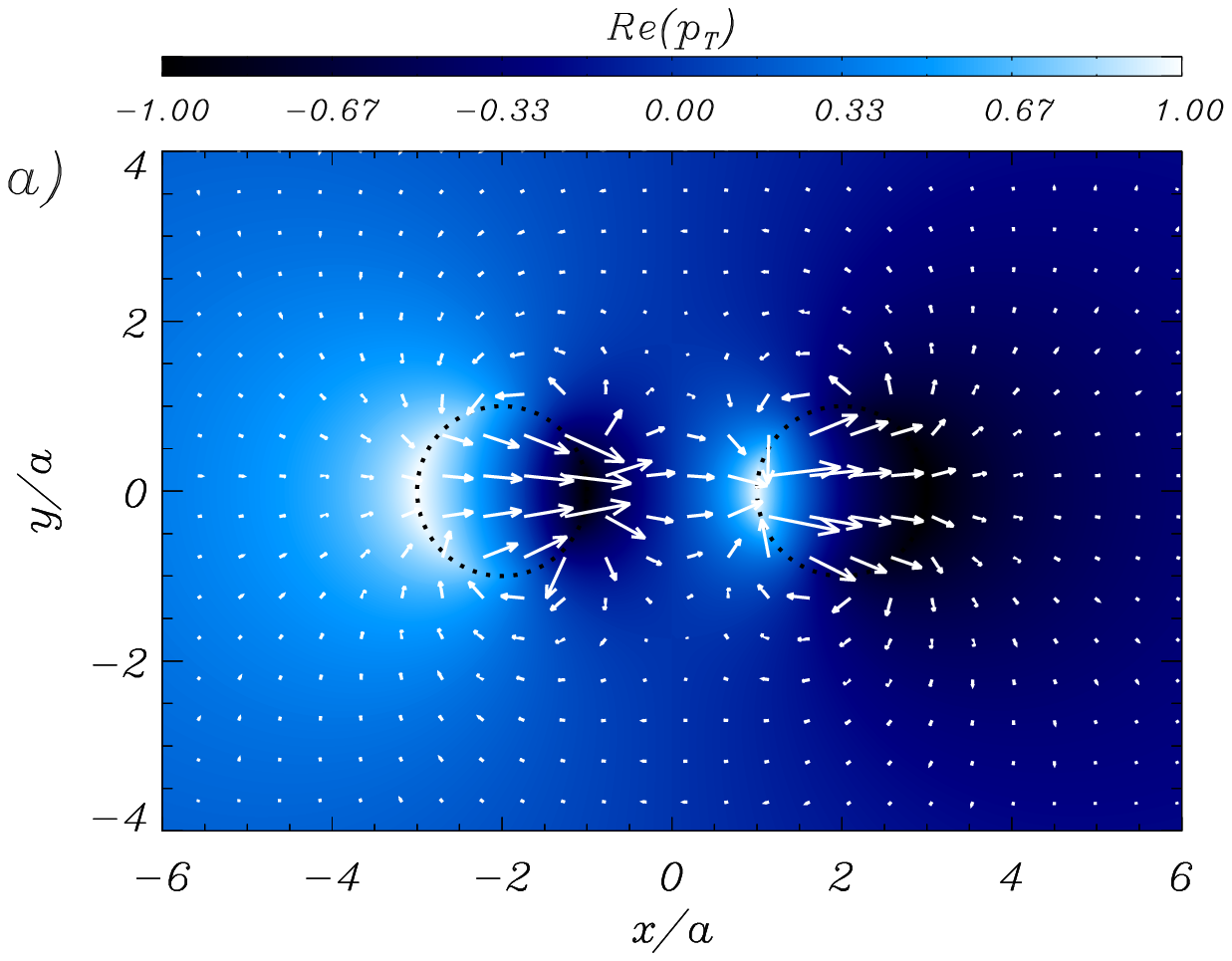}
\plotone{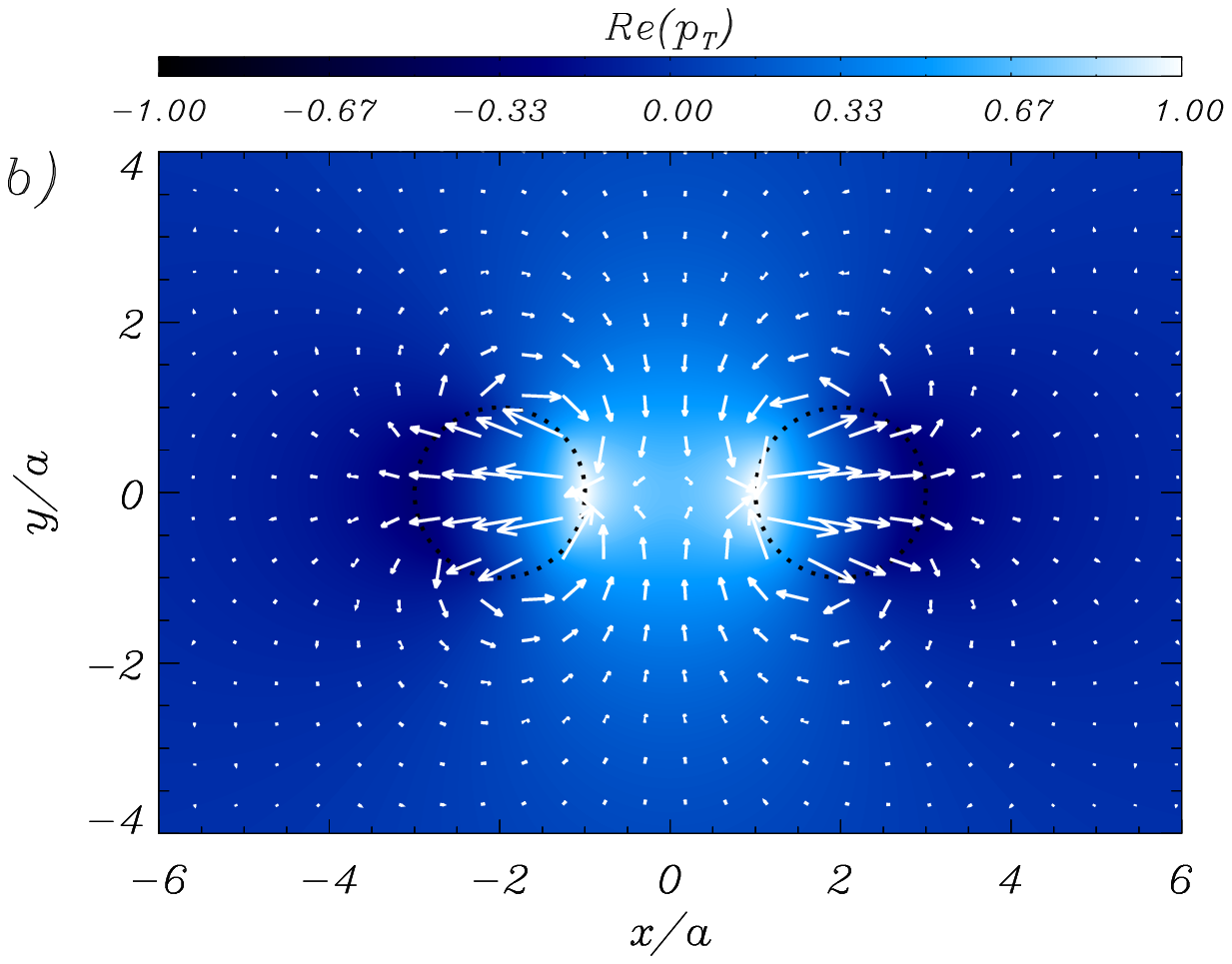}
\plotone{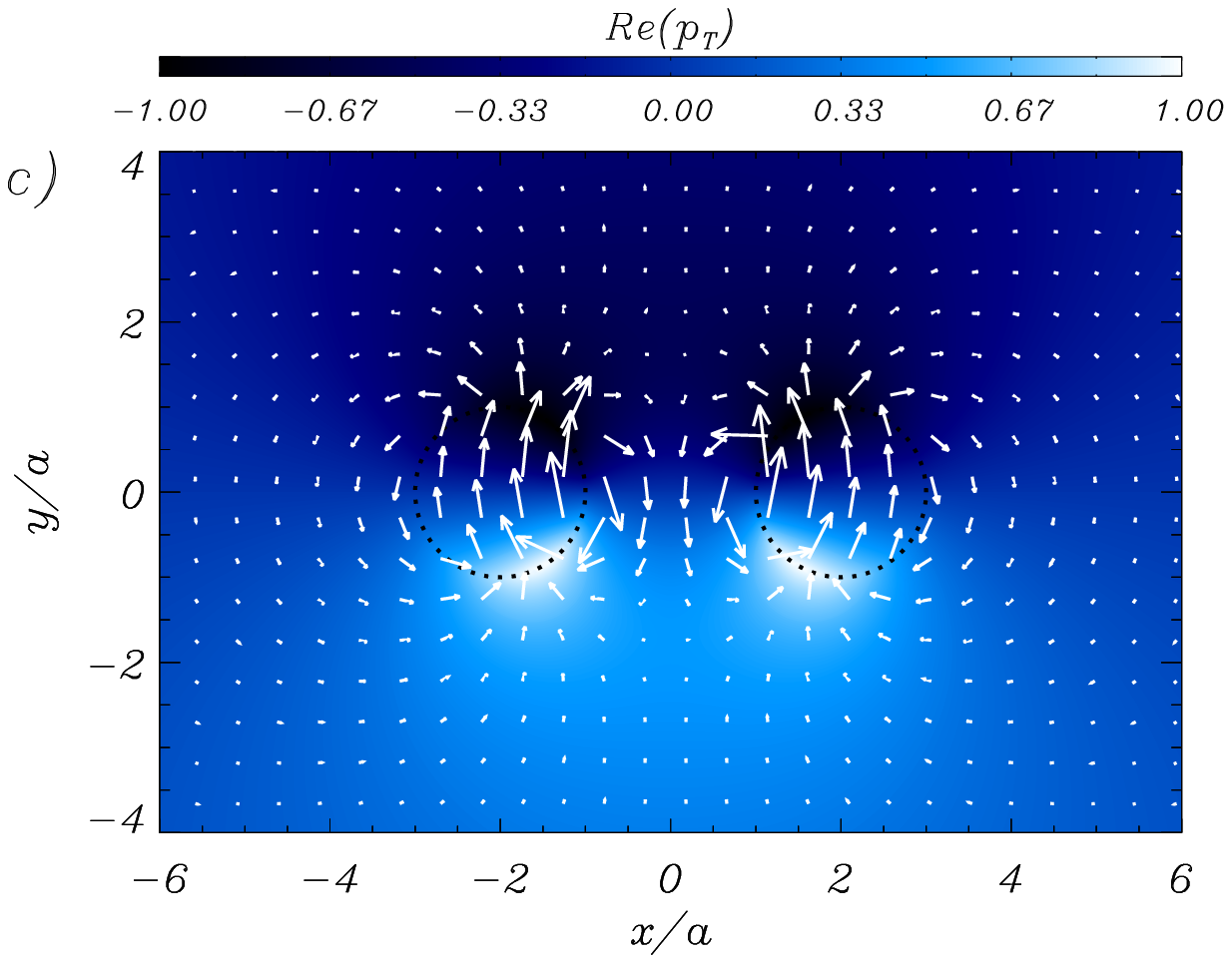}
\plotone{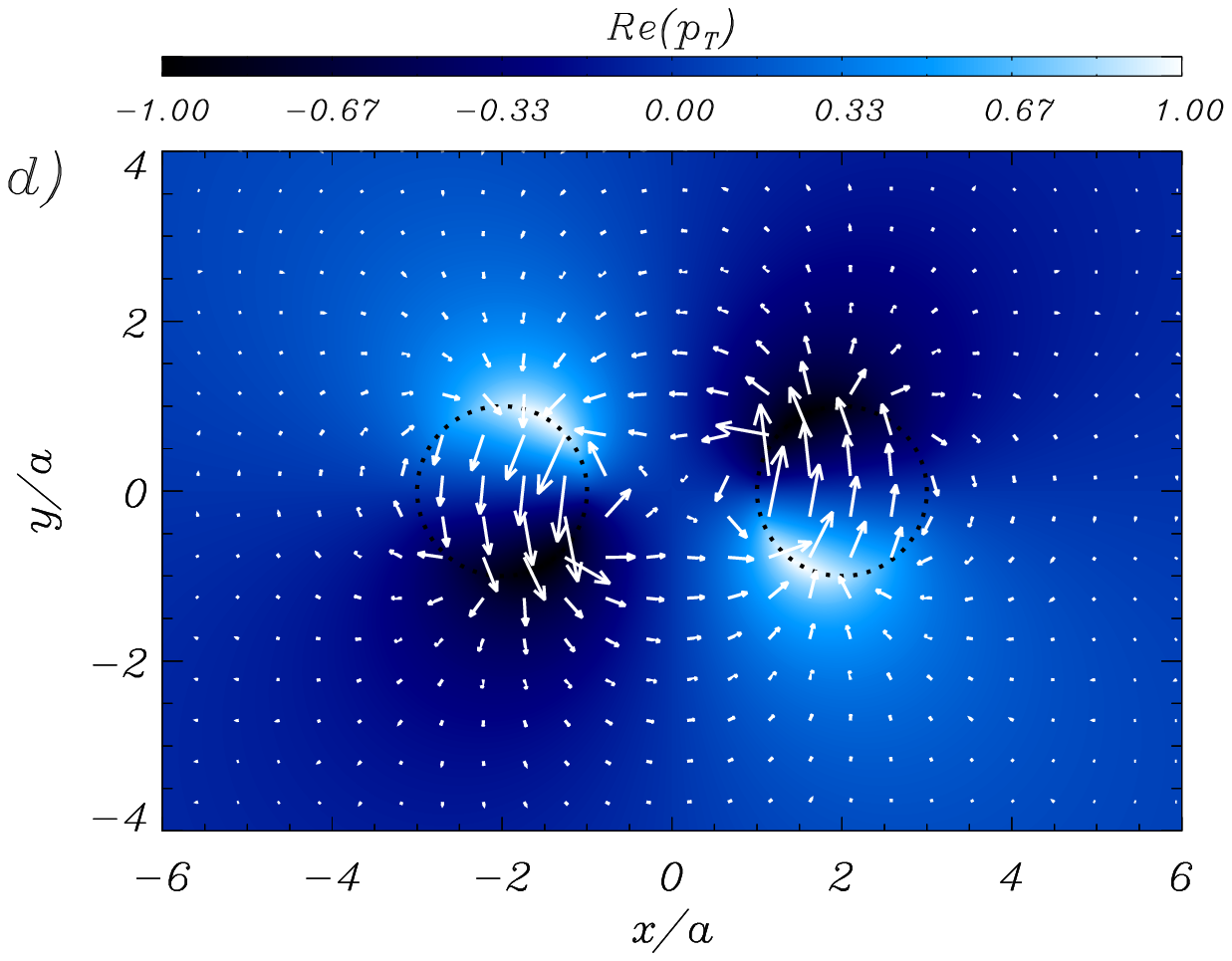}
\plotone{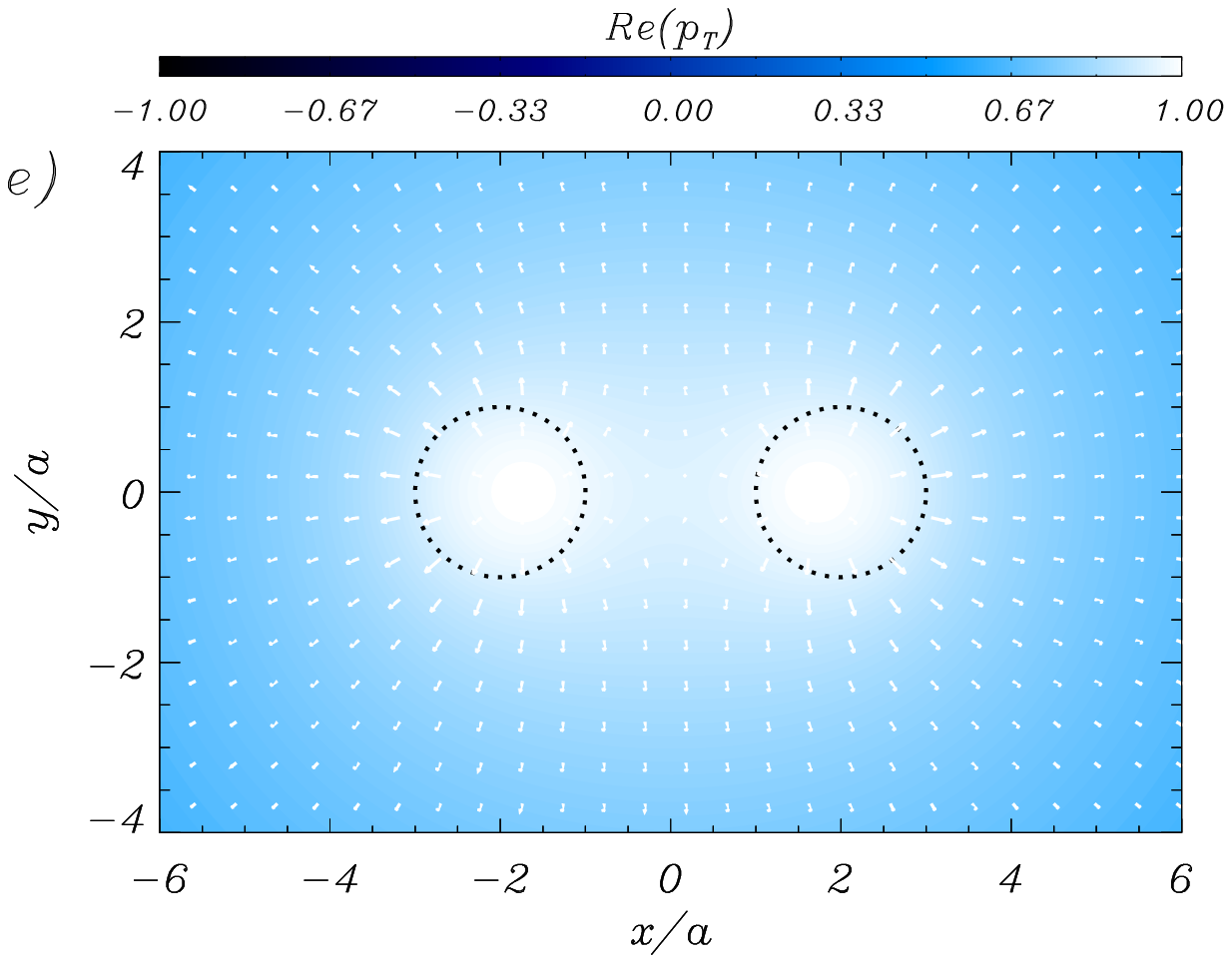}
\plotone{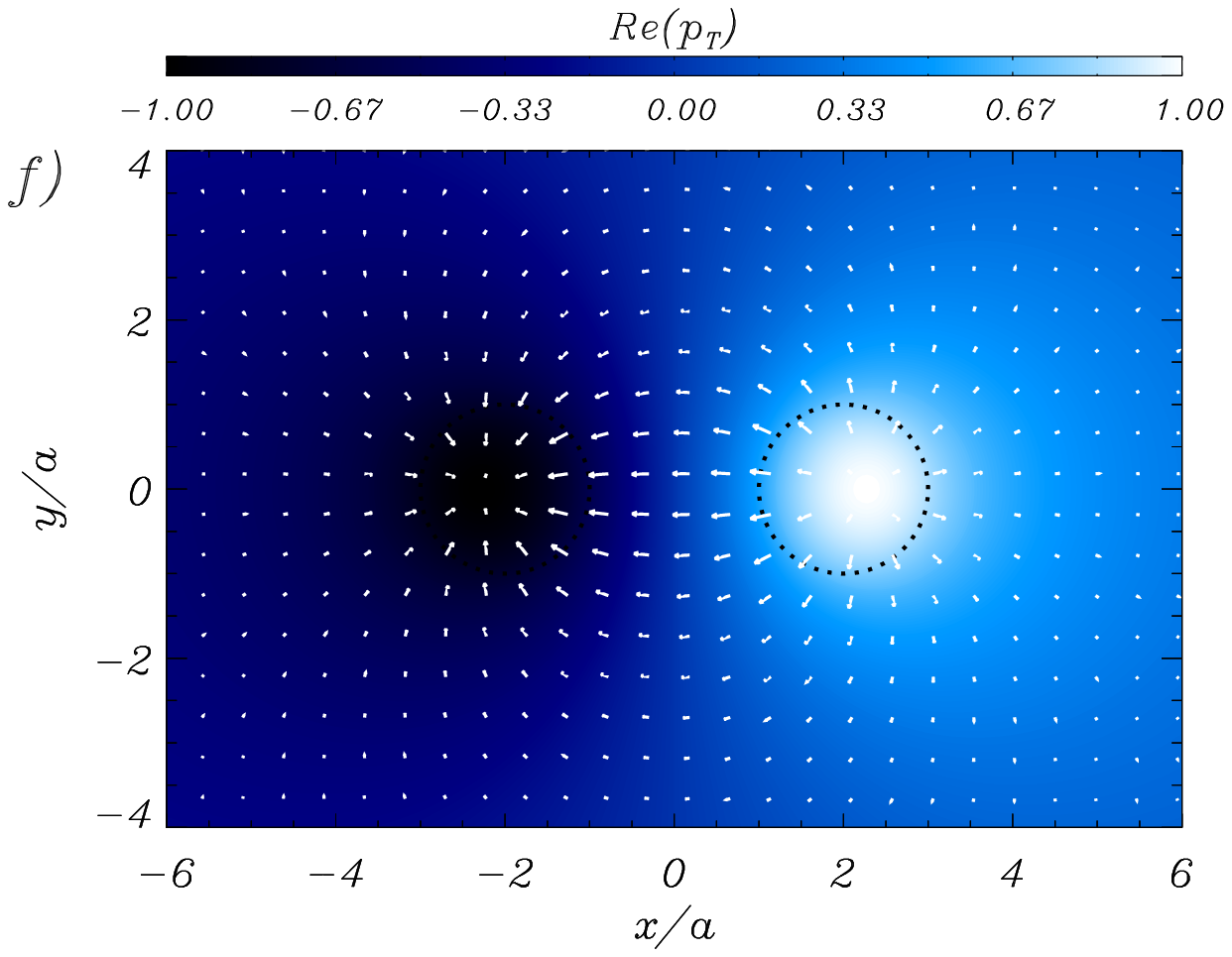}
% \plotone{f02a_bw.eps}
% \plotone{f02b_bw.eps}
% \plotone{f02c_bw.eps}
% \plotone{f02d_bw.eps}
% \plotone{f02e_bw.eps}
% \plotone{f02f_bw.eps}
\caption{Real part of the total pressure perturbation field (contour plot in arbitrary units) and the transverse Lagrangian displacement vector-field (arrows) plotted in the $xy$-plane corresponding to the wave modes $a)$~$S_x$, $b)$~$A_x$, $c)$~$S_y$, $d)$~$A_y$, $e)$~$S_z$, and $f)$~$A_z$ in the absence of flows for a separation between threads $d=4a$ and a longitudinal wavenumber $k_z a= 10^{-2}$. The location of prominence threads is denoted by dotted circles. \label{fig:eigen}}
\end{figure}

\begin{figure}[!htpb]
\centering
\epsscale{0.325}
\plotone{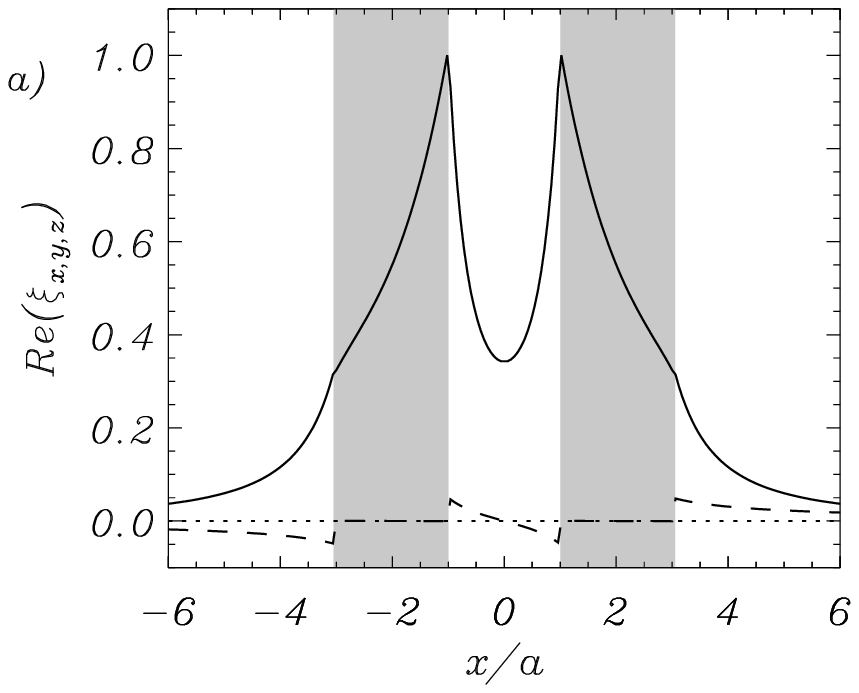}
\plotone{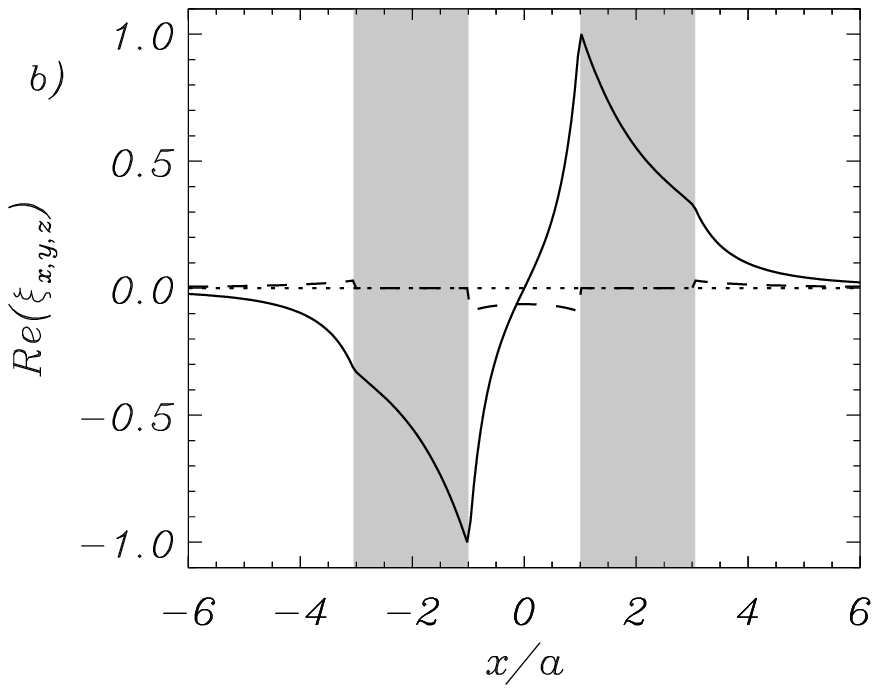}
\plotone{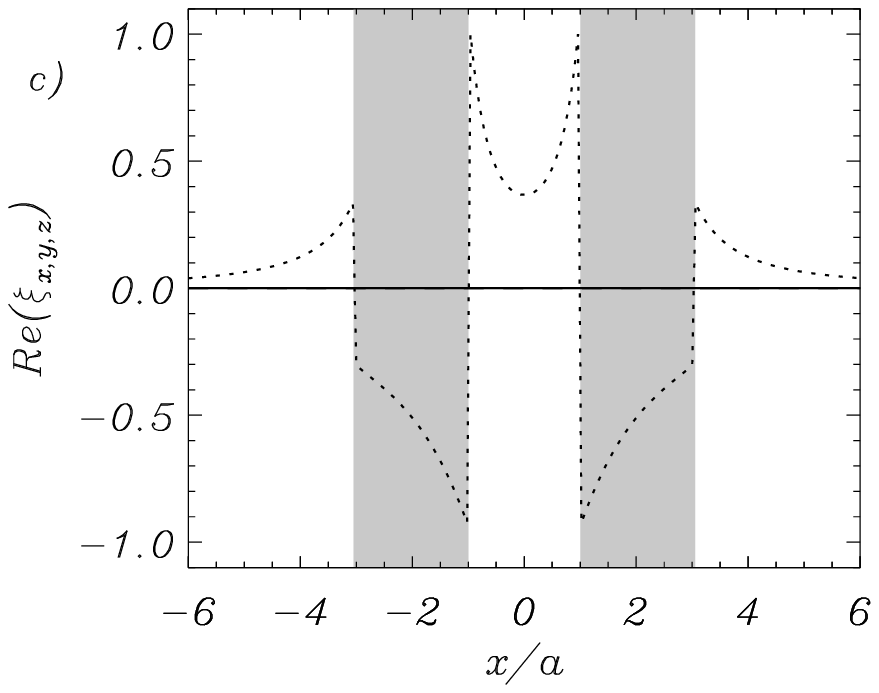}
\plotone{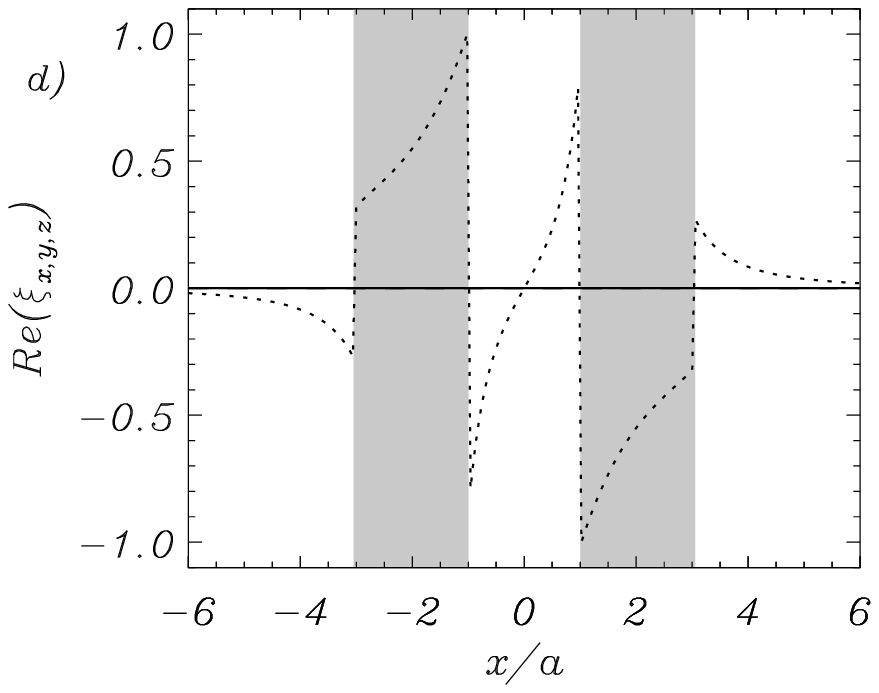}
\plotone{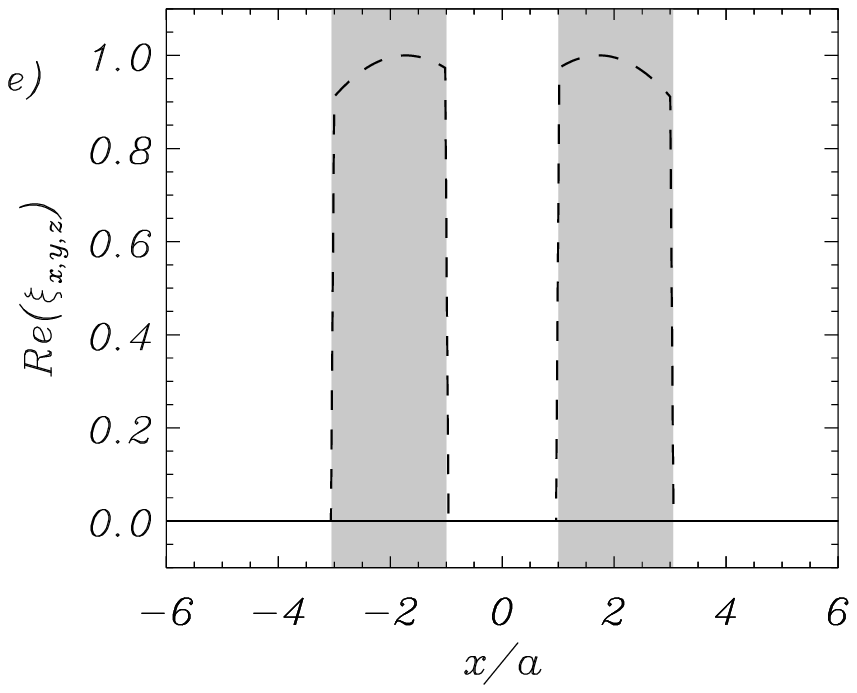}
\plotone{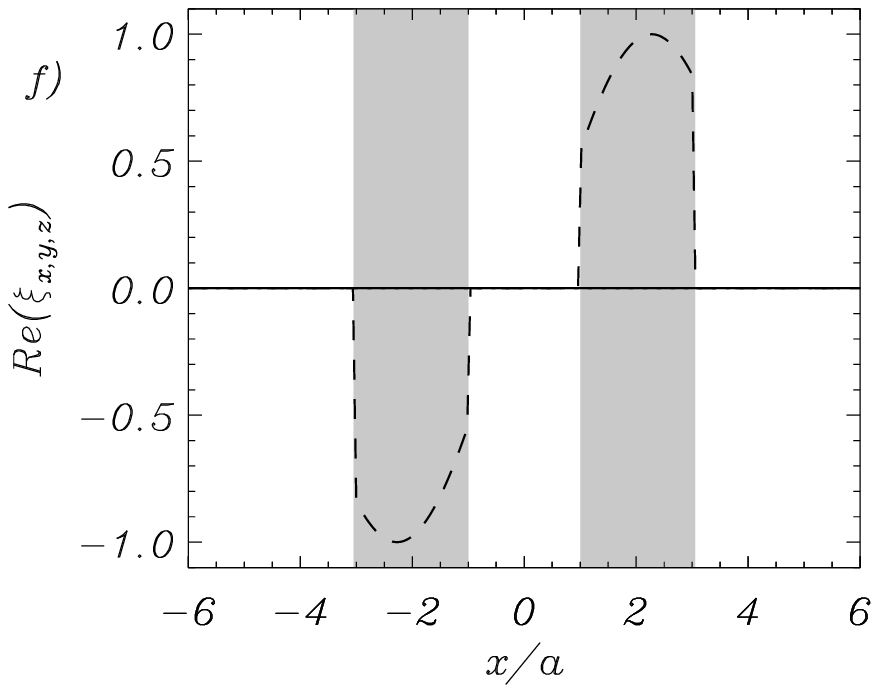}
\caption{Cut at $y=0$, $z=0$ of the real parts (in arbitrary units) of the Cartesian components of the Lagrangian displacement: $\xi_x$ (solid line), $\xi_y$ (dotted line), and $\xi_z$ (dashed line), corresponding to the wave modes $a)$~$S_x$, $b)$~$A_x$, $c)$~$S_y$, $d)$~$A_y$, $e)$~$S_z$, and $f)$~$A_z$ for the same conditions of Figure~\ref{fig:eigen}. The shaded regions show the location of threads. Note that neither $\xi_y$ nor $\xi_z$ are continuous at the edges of threads. \label{fig:cut}}
\end{figure}

The total pressure perturbation field, $p_{\rm T}$, and the transverse Lagrangian displacement vector-field, $\bf{\xi}_\perp$,  corresponding to the six fundamental modes are displayed in Figure~\ref{fig:eigen}.  On the other hand, Figure~\ref{fig:cut} displays a cut of the Cartesian components of the Lagrangian displacement ($\xi_x$, $\xi_y$, and $\xi_z$) at $y=0$, again for these six solutions. For simplicity, only the real part of these quantities are plotted in both Figures, since their imaginary parts are equivalent. One can see in Figure~\ref{fig:cut} that the amplitude of the longitudinal (magnetic field aligned) Lagrangian displacement, $\xi_z$, of the $S_z$ and $A_z$ modes is much larger than the amplitude of transverse displacements, $\xi_x$ and $\xi_y$, such as corresponds to slow modes in $\beta < 1$ homogeneous media, while the contrary occurs for the $S_x$, $A_x$, $S_y$, and $A_y$ fast kink solutions. 

\begin{figure}[!ht]
\centering
\epsscale{0.5}
\plotone{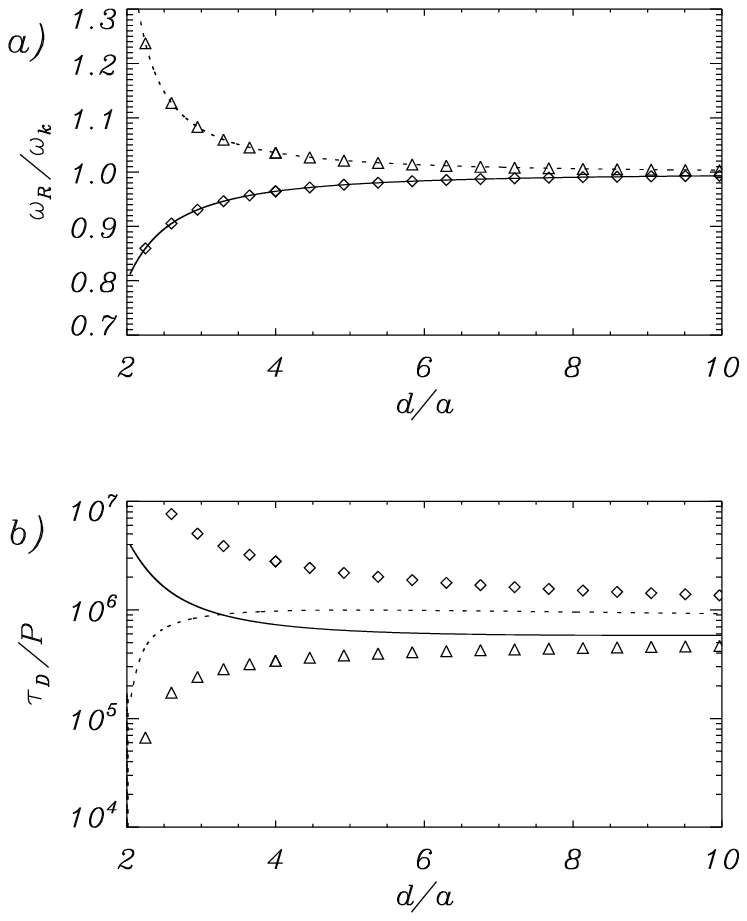}
\caption{$a)$ Ratio of the real part of frequency, $\omega_{\rm R}$, of the $S_x$ (solid line), $A_x$ (dotted line), $S_y$ (triangles), and $A_y$ (diamonds) wave modes to the frequency of the individual kink mode, $\omega_k$, as a function of the distance between centers. $b)$ Ratio of the damping time to the period versus the distance between centers. Linestyles are the same as in panel $a)$. \label{fig:distkink}}
\end{figure}

Next, Figure~\ref{fig:distkink}a displays the ratio of the real part of the frequency of the four kink solutions to the frequency of the individual kink mode, $\omega_k$ (from Paper~I), as a function of the distance between the center of cylinders, $d$. This Figure is equivalent to Figure~3 of \citet{luna1} and, in agreement with them, one can see that the smaller the distance between centers, the larger the interaction between threads and so the larger the separation between frequencies. On the other hand, Figure~\ref{fig:distkink}b shows the ratio of the damping time to the period of the four kink modes as a function of $d$. We see that the damping times are between 4 and 7 orders of magnitude larger than their corresponding periods. Therefore, dissipation by non-adiabatic mechanisms cannot be responsible for the observed damping times of transverse thread oscillations, as was pointed out in Paper~I. Recently, \citet{arregui} found that the mechanism of resonant absorption can provide kink mode damping times compatible with those observed.

\begin{figure}[!ht]
\centering
\epsscale{0.5}
\plotone{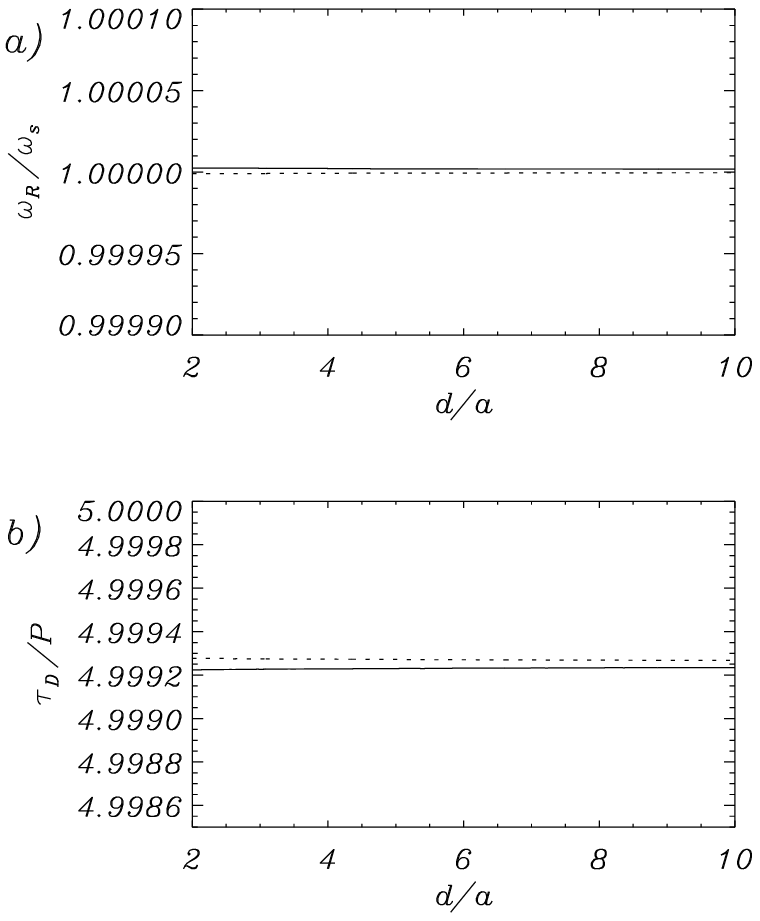}
\caption{$a)$ Ratio of the real part of frequency, $\omega_{\rm R}$, of the $S_z$ (solid line) and $A_z$ (dotted line) wave modes to the frequency of the individual slow mode, $\omega_s$, as a function of the distance between centers. $b)$ Ratio of the damping time to the period versus the distance between centers. Linestyles are the same as in panel $a)$. \label{fig:distslow}}
\end{figure}

Regarding slow modes, Figure~\ref{fig:distslow}a displays the ratio of the real part of the frequency of the $S_z$ and $A_z$ solutions to the frequency of the individual slow mode, $\omega_s$ (from Paper~I). One can see that the frequencies of the $S_z$ and $A_z$ modes are almost identical to the individual slow mode frequency, and so the strength of the interaction is almost independent of the distance between cylinders. This is consistent with the fact that transverse motions (responsible for the interaction between threads) are not significant for slow-like modes in comparison with their longitudinal motions. Therefore, the $S_z$ and $A_z$ modes essentially behave as individual slow modes, contrary to kink modes, which display a more significant collective behavior. Finally, Figure~\ref{fig:distslow}b shows $\td / P$ corresponding to the $S_z$ and $A_z$ solutions versus $d$. One sees that both slow modes are efficiently attenuated by non-adiabatic mechanisms, with $\td / P \approx 5$, which is in agreement with previous studies \citep[][]{soler1, solerapj} and consistent with observations.

\subsubsection{Effect of steady mass flows on the collective behavior of wave modes}

The aim of the present section is to assess the effect of flows on the behavior of collective modes.  With no loss of generality, we assume no flow in the corona, i.e. $U_{\rm c} = 0$. On the other hand, the flow velocities in both cylinders, namely $U_1$ and $U_2$, are free parameters. We vary these flow velocities between -30~km~s$^{-1}$ and 30~km~s$^{-1}$, which correspond to the range of typically observed flow velocities in filament threads \citep[e.g.,][]{lin2003}. These flow velocities are below the critical value that determines the apparition of the Kelvin-Helmholtz instability \citep[see details in][]{holzwarth}. In our configuration, a positive flow velocity means that the mass is flowing towards the positive $z$-direction, whereas the contrary is for negative flow velocities. From Paper~I \citep[see also][]{terra} we know that the symmetry between waves whose propagation is parallel ($\omega_{\rm R} > 0$) or anti-parallel ($\omega_{\rm R} < 0$) with respect to magnetic field lines is broken by the presence of flows. Hence, we must take into account the direction of wave propagation in order to perform a correct description of the wave behavior. Following Paper~I, we call parallel waves those solutions with $\omega_{\rm R} > 0$, while anti-parallel waves are solutions with $\omega_{\rm R} < 0$.

\begin{figure}[!ht]
\centering
\epsscale{0.65}
\plotone{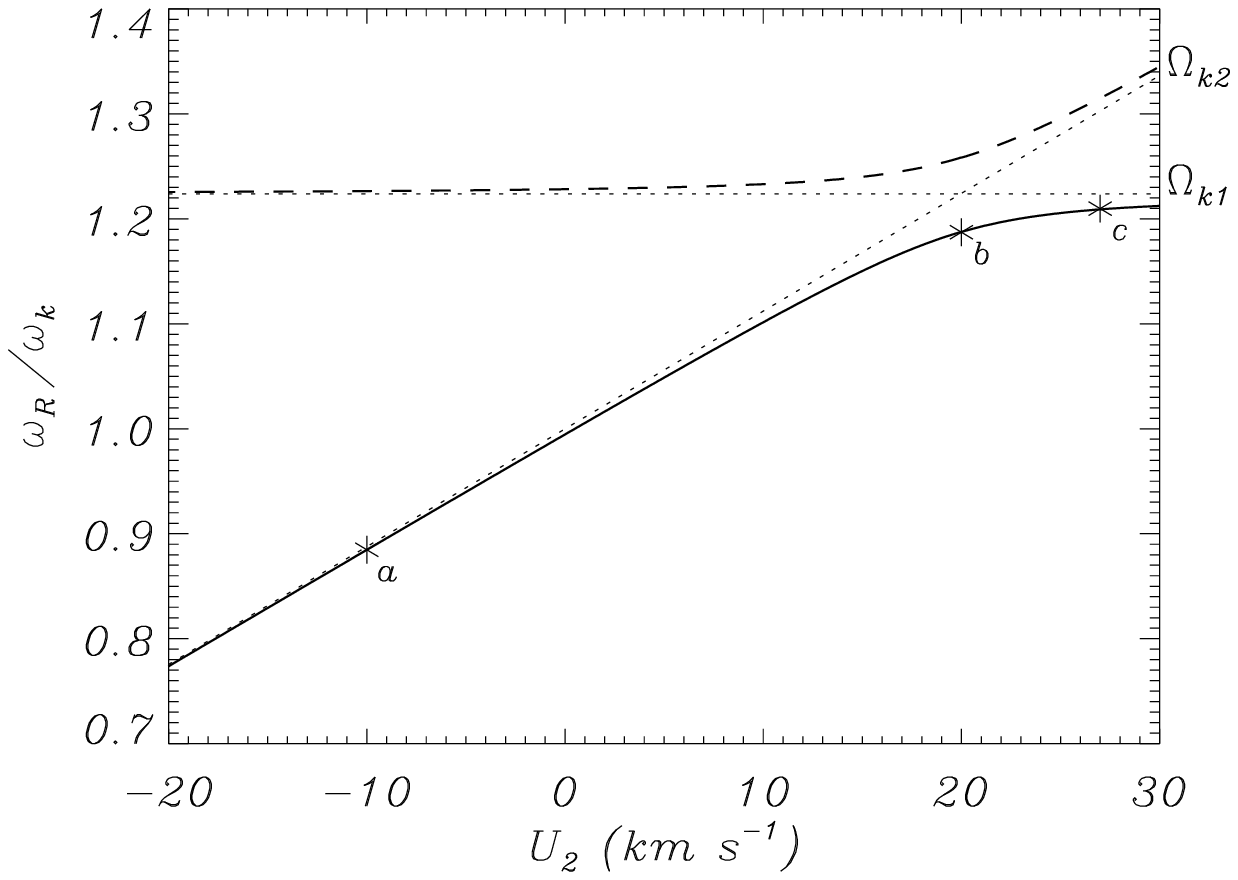}
\caption{Ratio of the real part of the frequency, $\omega_{\rm R}$, to the individual kink frequency, $\omega_k$, as a function of $U_2$ for $U_1=$~20~km~s$^{-1}$. The solid line corresponds to parallel low-frequency modes ($S_x$ and $A_y$) while the dashed line corresponds to parallel high-frequency solutions ($A_x$ and $S_y$). Dotted lines correspond to the Doppler-shifted individual kink frequencies of the threads,  $\Omega_{k 1}$ and $\Omega_{k 2}$. The small letters next to the solid line refer to particular situations studied in the text. \label{fig:phase}}
\end{figure}

We begin this investigation with transverse modes. First, we assume $U_1 =$~20~km~s$^{-1}$ and study the behavior of the oscillatory frequency when $U_2$ varies (see Fig.~\ref{fig:phase}). Since frequencies are almost degenerate and, therefore, almost indiscernible if they are plotted together, we use the notation of \citet{tom} and call low-frequency modes the $S_x$ and $A_y$ solutions, while high-frequency modes refer to $A_x$ and $S_y$ solutions. In addition, we restrict ourselves to parallel propagation because the argumentation can be easily extended to anti-parallel waves. To understand the asymptotic behavior of frequencies in Figure~\ref{fig:phase}, we define the following Doppler-shifted individual kink frequencies:
\begin{equation}
  \Omega_{k 1} = \omega_k + U_1 k_z, \label{eq:wkleft}
\end{equation}
\begin{equation}
  \Omega_{k 2} = \omega_k + U_2 k_z. \label{eq:wkright}
\end{equation}
Since $U_1$ is fixed, $\Omega_{k 1}$ is a horizontal line in Figure~\ref{fig:phase}, whereas $\Omega_{k 2}$ is linear with $U_2$.

\begin{figure}[!htbp]
\centering
\epsscale{0.32}
\plotone{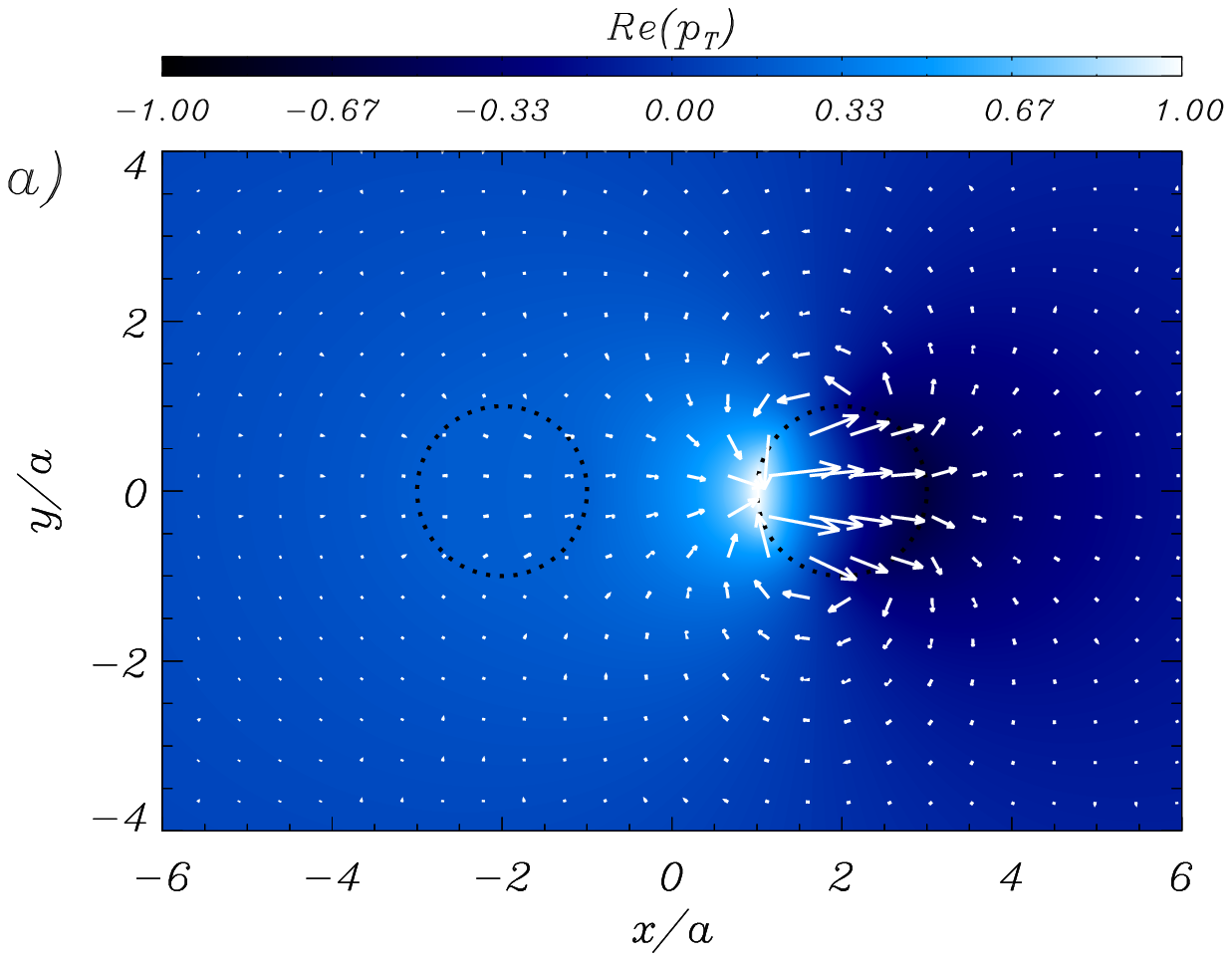}
\plotone{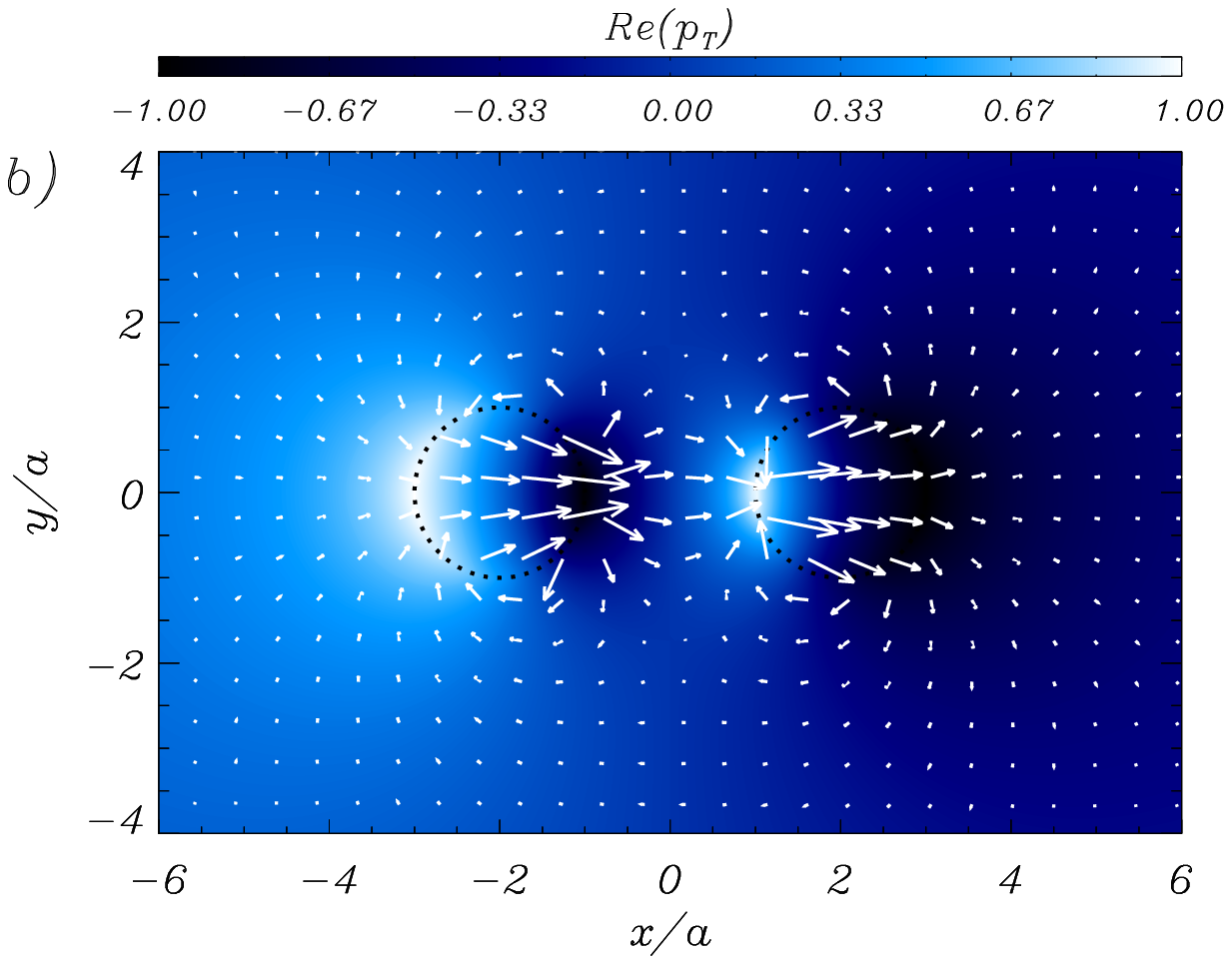}
\plotone{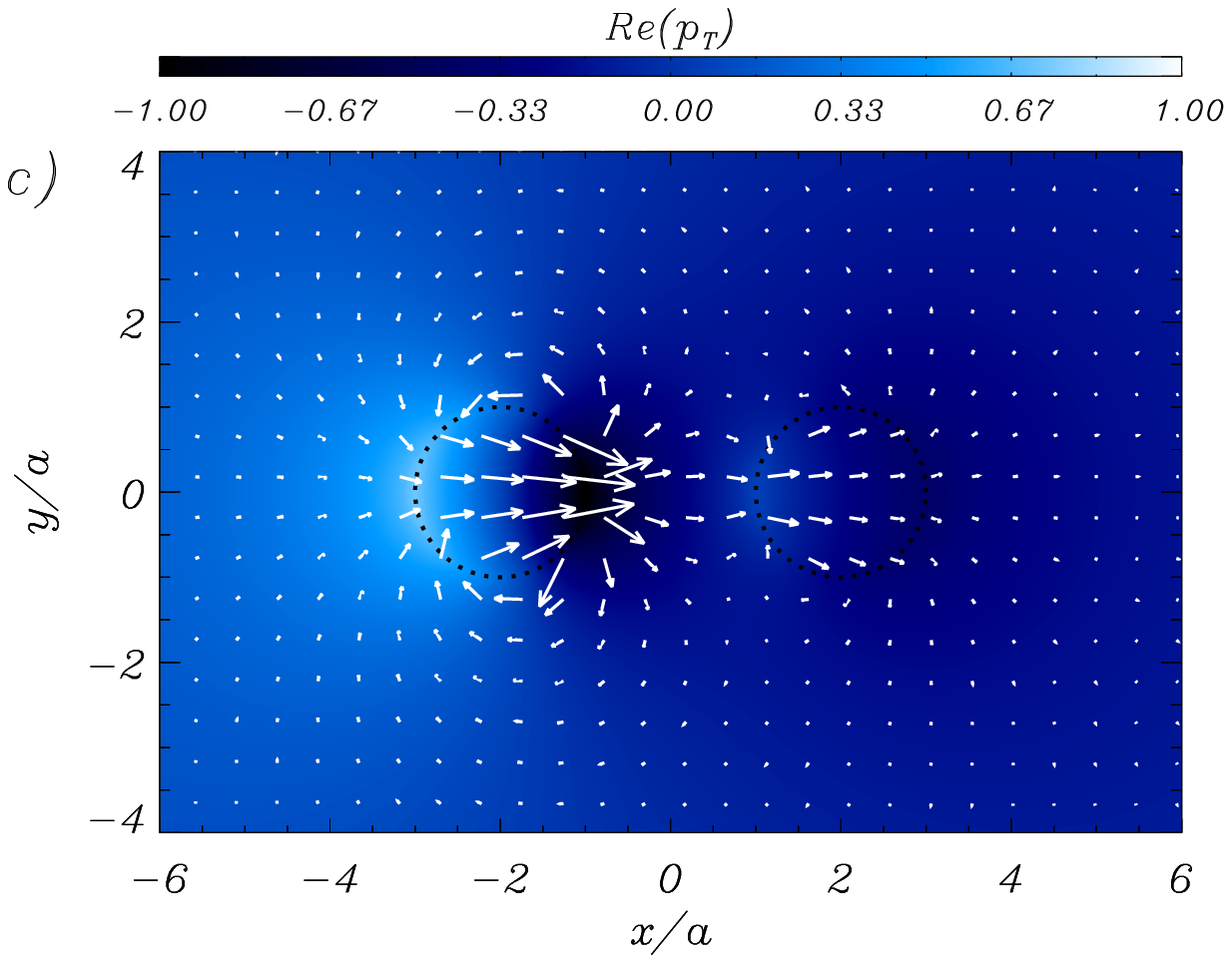}
% \plotone{f07a_bw.eps}
% \plotone{f07b_bw.eps}
% \plotone{f07c_bw.eps}
\caption{Real part of the total pressure perturbation field (contour plot in arbitrary units) and the transverse Lagrangian displacement vector-field (arrows) plotted in the $xy$-plane corresponding to the parallel $S_x$ mode for $a)$ $U_2 =$~-10~km~s$^{-1}$, $b)$ $U_2 =$~20~km~s$^{-1}$, and $c)$ $U_2 =$~27~km~s$^{-1}$. In all cases $U_1 =$~20~km~s$^{-1}$. \label{fig:eigenampl}}
\end{figure}

 Three interesting situations have been pointed by means of small letters from $a$ to $c$ in Figure~\ref{fig:phase}. Each of these letters also corresponds to a panel of Figure~\ref{fig:eigenampl} in which the total pressure perturbation field of the $S_x$ mode is plotted. The three different situations are commented in detail next (remember that in all cases $U_1 = 20$~km~s$^{-1}$):

\begin{itemize}
 \item $a)$ $U_2 = -10$~km~s$^{-1}$ ($U_2 < U_1$). This corresponds to a situation of counter-streaming flows. From Figure~\ref{fig:phase} we see that the frequency of low-frequency modes is close to $\Omega_{k 2}$, whereas that of high-frequency solutions is near $\Omega_{k 1}$. Thus, these solutions do not interact with each other and low-frequency (high-frequency) solutions are related to individual oscillations of the second (first) thread. This is verified by looking at the total pressure perturbation field in Figure~\ref{fig:eigenampl}a, that shows that only the second thread is significantly perturbed. Therefore, for an external observer this situation corresponds in practice to an individual thread oscillation.
\item $b)$ $U_2 = 20$~km~s$^{-1}$ ($U_2 = U_1$). The flow velocities and their directions are equal in both threads. In such a situation, low- and high-frequency modes couple. At the coupling, an avoided crossing of the solid and dashed lines is seen in Figure~\ref{fig:phase}. Because of this coupling solutions are related no more to oscillations of an individual thread but they are now collective and, for this reason, Figure~\ref{fig:eigenampl}b shows a significant pressure perturbation in both threads.
\item $c)$ $U_2 = 27$~km~s$^{-1}$ ($U_2 > U_1$). This case is the opposite to situation $a)$. Therefore, in practice the present situation corresponds again to an individual thread oscillation.
\end{itemize}

\begin{figure}[!ht]
\centering
\epsscale{0.65}
\plotone{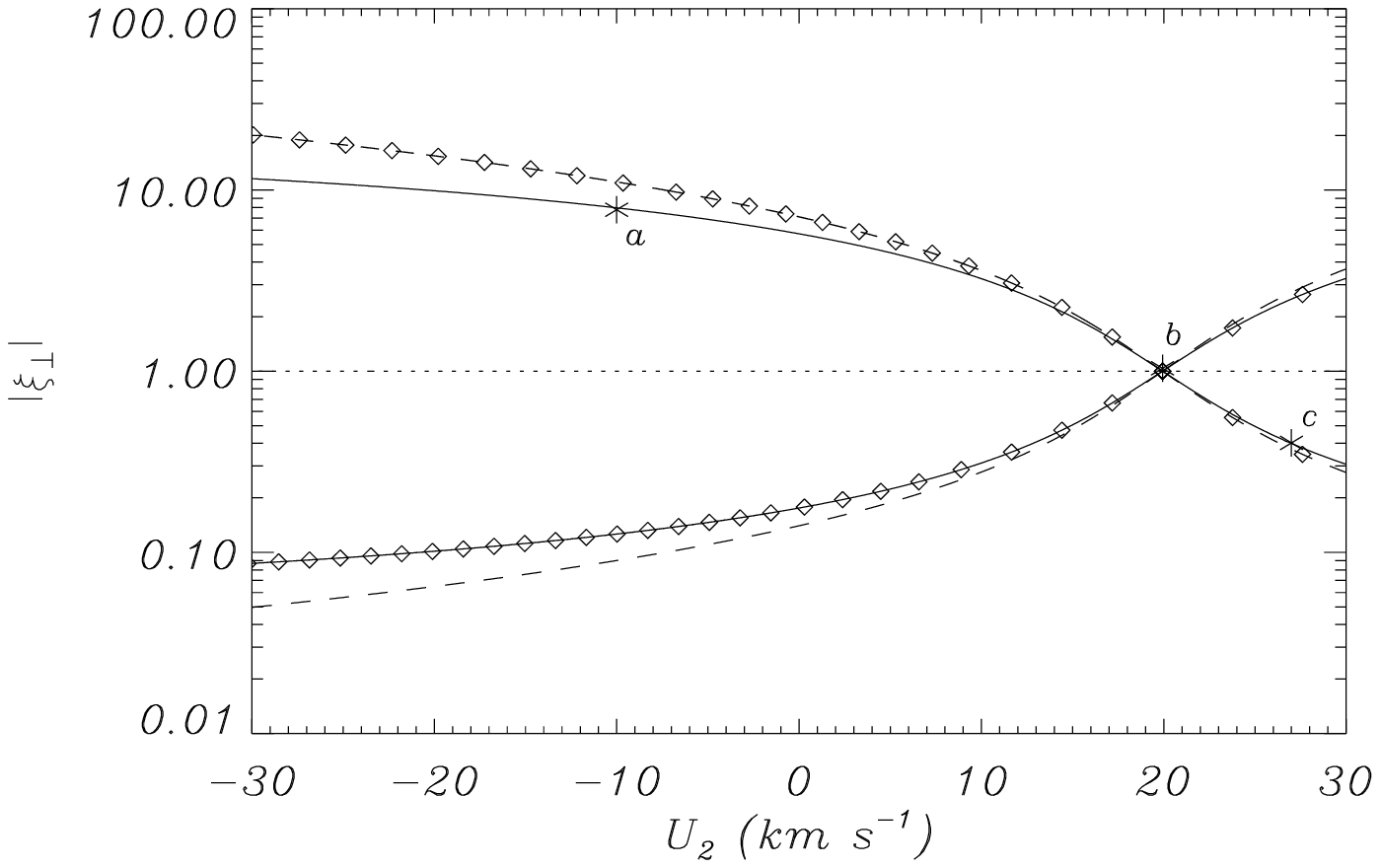}
\caption{Amplitude (in arbitrary units) of the transverse Lagrangian displacement, $\xi_\perp$, at the center of the second thread as a function of $U_2$. The amplitude displacement in the first thread center is fixed to unity (dotted line) and the flow velocity is $U_{\rm l} =$~20~km~s$^{-1}$. Solid lines correspond to low-frequency kink modes ($S_x$ and $A_y$), while dashed lines correspond to high-frequency kink solutions ($A_x$ and $S_y$). Lines without simbols are for parallel waves whereas diamonds indicate anti-parallel propagation. Small letters refer to the particular situations discussed in the text. \label{fig:ampli}}
\end{figure}

Our argumentation is supported by Figure~\ref{fig:ampli}, which displays the amplitude of the transverse Lagrangian displacement, $\xi_\perp$, at the center of the second thread as a function of $U_2$, for parallel and anti-parallel kinklike waves. The displacement amplitude at the center of the first thread is always set equal to unity. The three previously commented situations have been pointed again in Figure~\ref{fig:ampli}. We clearly see that the displacement amplitude is only comparable in both threads, and so their dynamics is collective, when their flow velocities are similar.

\begin{figure}[!htbp]
\centering
\epsscale{0.4}
\plotone{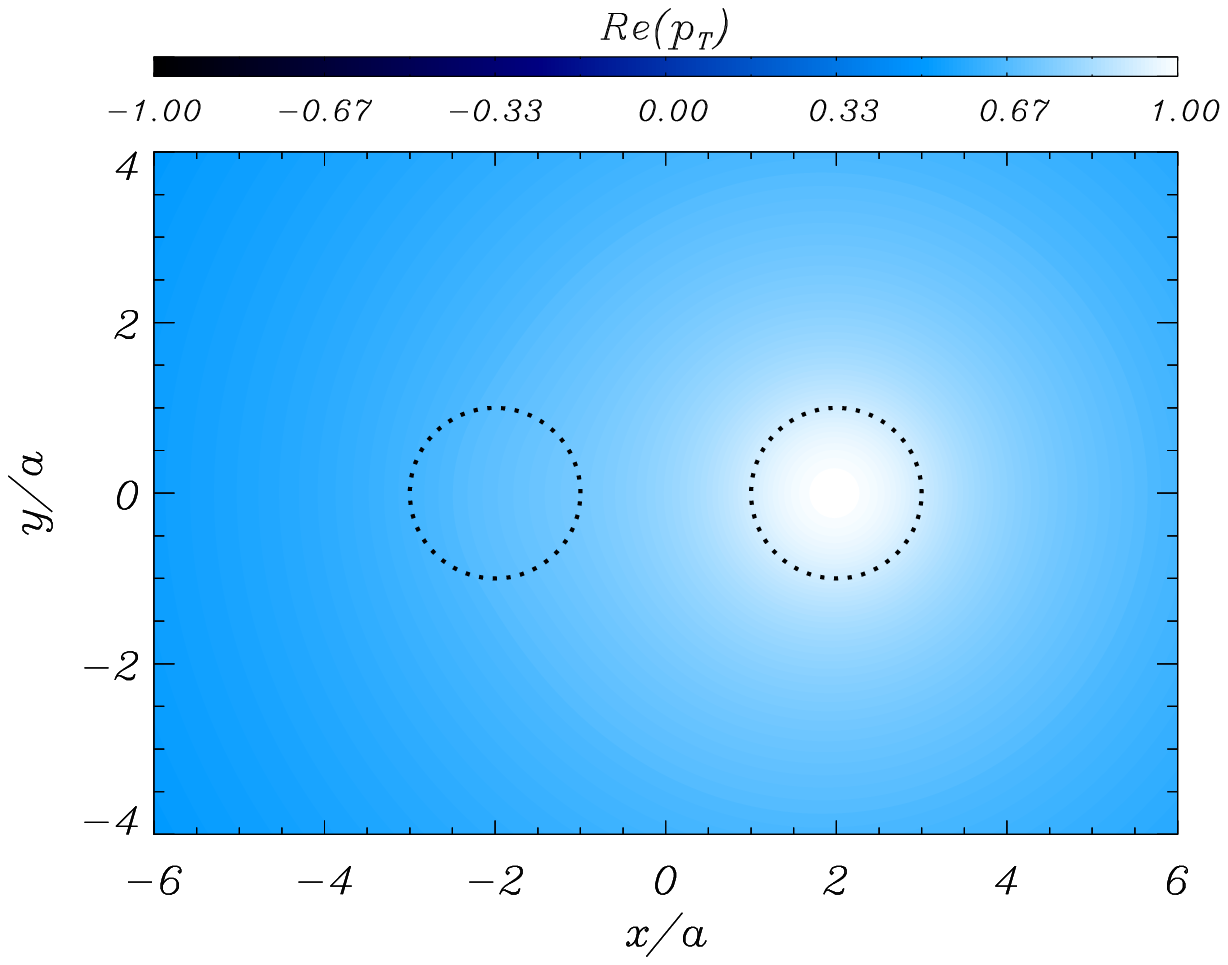}
%\plotone{f09_bw.eps}
\caption{Real part of the total pressure perturbation field plotted in the $xy$-plane corresponding to the parallel $A_z$ slow mode for a difference between flow velocities of $U_1 - U_2=10^{-3}$~km~s$^{-1}$. \label{fig:eigenamplslow}}
\end{figure}

Next we turn our attention to slow modes. The behavior of the $S_z$ and $A_z$ modes is like that of low- and high-frequency kinklike solutions, so we comment them in short for the sake of simplicity.  $S_z$ and $A_z$ solutions can only be considered collective when the flow velocity is the same in both threads because, in such a case, the $S_z$ and $A_z$ modes couple. If different flows within the threads are considered, the $S_z$ and $A_z$ slow modes lose their collective aspect and their frequencies are very close to the Doppler-shifted individual slow frequencies,
\begin{equation}
  \Omega_{s 1} = \omega_s + U_1 k_z, \label{eq:wsleft}
\end{equation}
\begin{equation}
  \Omega_{s 2} = \omega_s + U_2 k_z. \label{eq:wsright}
\end{equation}
Then, the $S_z$ and $A_z$ solutions behave like individual slow modes. It is worth to mention that the coupling between slow modes is much more sensible to the flow velocities in comparison with fast modes, and the $S_z$ and $A_z$ solutions quickly decouple if  $U_1$ and $U_2$ slightly differ. An example of this behavior is seen in Figure~\ref{fig:eigenamplslow}, which displays the total pressure perturbation field of the $A_z$ mode for $U_1 - U_2=10^{-3}$~km~s$^{-1}$. Although the difference of the flow velocities is insignificant, one can see that in this situation the $A_z$ mode essentially behaves as the individual slow mode of the second thread.

The main idea behind these results is that fast or slow wave modes with a collective appearance (i.e., modes with a similar displacement amplitude within all threads) are only possible when the Doppler-shifted individual kink (eqs.~[\ref{eq:wkleft}] and [\ref{eq:wkright}]) or slow (eqs.~[\ref{eq:wsleft}] and [\ref{eq:wsright}]) frequencies are similar in both threads. In a system of identical threads, this can only be achieved by considering the same flow velocities within all threads, since all of them have the same individual kink and slow frequencies. However, if threads with different physical properties are considered (i.e., different individual frequencies), it is possible that the coupling may occur for different flow velocities. This is explored in the next section.

\subsection{Configuration of two threads with different physical properties}
\label{sec:ntreads}

Now we consider a system of two nonidentical threads and focus first on kink modes. From \S~\ref{sec:2treads} we expect that collective kink motions occur when the Doppler-shifted individual kink frequencies of both threads coincide. The relation between flow velocities $U_1$ and $U_2$ for which the coupling takes place can be easily estimated from the equation,
\begin{equation}
\omega_{k 1} + U_1 k_z \approx \omega_{k 2} + U_2 k_z.
\end{equation}
Note that the individual kink frequency of each thread is different, i.e., $\omega_{k 1} \neq \omega_{k 2}$. Then, the relation between flow velocities at the coupling is,
\begin{equation}
 U_1 - U_2  \approx \frac{\omega_{k 2} - \omega_{k 1}}{k_z}. 
\end{equation}
This last expression can be simplified by considering the approximate expression for the kink frequency in the long-wavelength limit,
\begin{equation}
 \omega_{k i} \approx \pm \sqrt{\frac{2}{1+\rho_c / \rho_i}} v_{{\rm A} i} k_z \approx \pm \sqrt{2}\, v_{{\rm A} i} k_z,
\end{equation}
for $i=1,2$, where the $+$ sign is for parallel waves and the $-$ sign is for anti-parallel propagation. Then, one finally obtains,
\begin{equation}
 U_1 - U_2  \approx \pm \sqrt{2} \left( v_{{\rm A} 2} - v_{{\rm A} 1} \right), \label{eq:velrelation}
\end{equation}
where the meaning of the $+$ and $-$ signs is the same as before. In the case of identical threads, $v_{{\rm A} 1} = v_{{\rm A} 2}$ and so  $U_1 - U_2  = 0$. Thus the flow velocity must be the same in both threads to obtain collective motions, as we concluded in \S~\ref{sec:2treads}. An equivalent analysis can be performed for slow modes and one obtains,
\begin{equation}
   U_1 - U_2  \approx \pm \left( c_{s 2} - c_{s 1} \right). \label{eq:velrelationslow}
\end{equation}
In general,  the coupling between slow modes occur at different flow velocities than the coupling between kink modes. This makes difficult the simultaneous existence of collective slow and kink solutions in systems of non-identical threads.

\begin{figure}[!h]
\centering
\epsscale{0.65}
\plotone{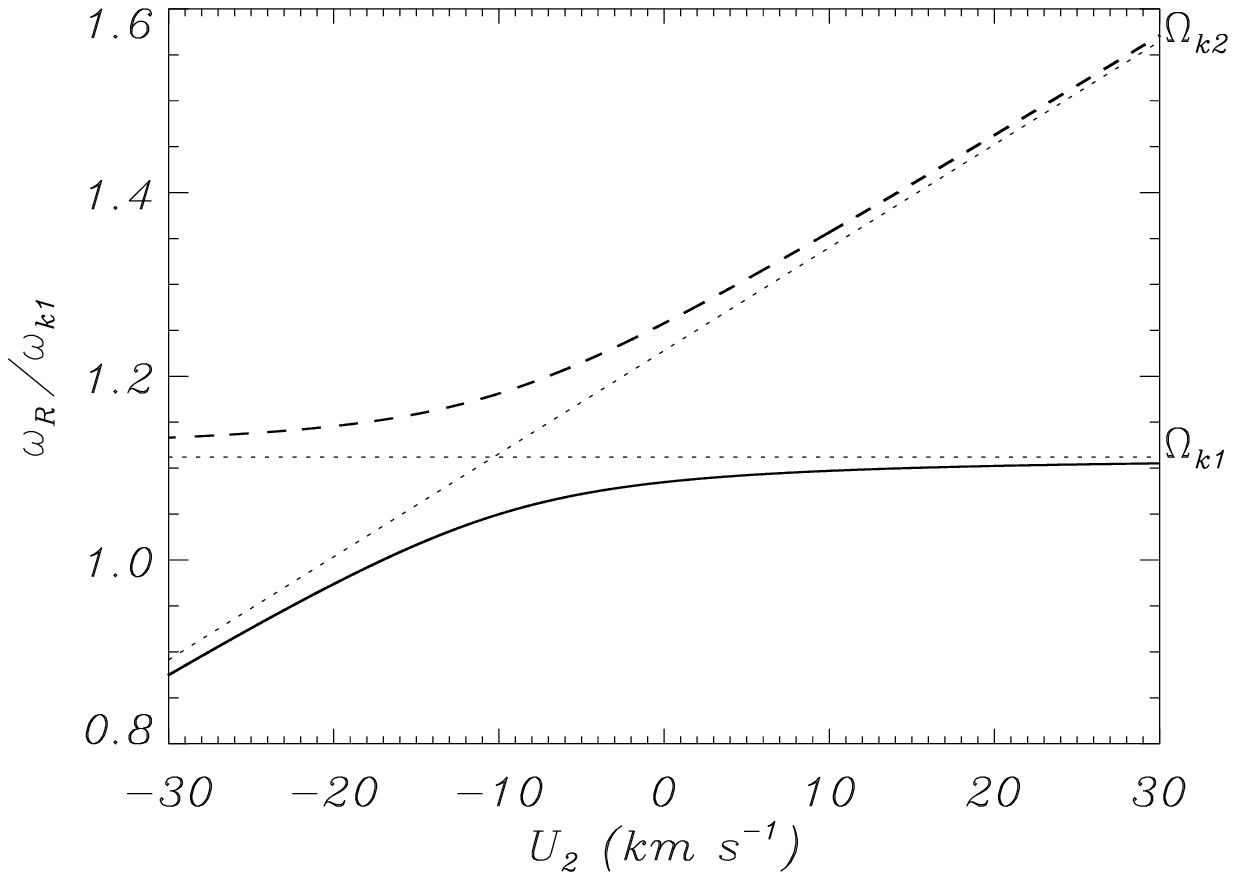}
\caption{Same as Figure~\ref{fig:phase} but for the particular configuration studied in \S~\ref{sec:ntreads}. \label{fig:phasediff}}
\end{figure}

Next, we assume a particular configuration of two non-identical threads in order to verify the later argumentation. Thread radii are $a_1=30$~km and $a_2 = 45$~km, whereas their physical properties are $T_1=8000$~K, $\rho_1 = 5 \times 10^{-11}$~kg~m$^{-3}$, and $T_2=12000$~K, $\rho_2 = 3.33 \times 10^{-11}$~kg~m$^{-3}$. Coronal conditions are those taken in \S~\ref{sec:2treads} (i.e., $T_{\rm c} = 10^6$~K, $\rho_{\rm c} = 2.5 \times 10^{-13}$~kg~m$^{-3}$). The magnetic field strength is 5~G everywhere and the distance between the thread centers is $d= 120$~km. We assume $U_1 = 10$~km~s$^{-1}$. In the present case, also four kinklike solutions are present, which are grouped in two almost degenerate couples. For this reason, we again refer to them as low- and high-frequency kink solutions. Their frequency as a function of $U_2$ is displayed in Figure~\ref{fig:phasediff}. At first sight, we see that solutions couple for a particular value of $U_2$, as expected. Applying equation~(\ref{eq:velrelation}), and considering that $v_{{\rm A} 1} = 63.08$~km~s$^{-1}$ and $v_{{\rm A} 2} = 77.29$~km~s$^{-1}$, we obtain $ U_1- U_2  \approx \pm 20.10\,\rm{km}\,\rm{s}^{-1}$, and since $U_1=10$~km~s$^{-1}$, we get $U_2 \approx - 10.10$~km~s$^{-1}$ for parallel propagation. We see that the approximate value of $U_2$ obtained from equation~(\ref{eq:velrelation}) is in good agreement with Figure~\ref{fig:phasediff}. In addition, we obtain that for parallel propagation collective dynamics appear in a situation of counter-streaming (opposite) flows. This result is of special relevance because counter-streaming flows have been detected in prominences \citep{zirker, lin2003} and might play a crucial role in the collective behavior of oscillations. On the contrary, in the anti-parallel propagation case we obtain $U_2 \approx 30.10$~km~s$^{-1}$ from equation~(\ref{eq:velrelation}), and so both flows are in the same direction and large a quite value of  $U_2$ is obtained in comparison with the parallel propagation case. 

Regarding slow modes, considering that $c_{s 1} = 11.76$~km~s$^{-1}$ and $c_{s 2} = 14.40$~km~s$^{-1}$, equation~(\ref{eq:velrelationslow}) gives $U_2 \approx 7.36$~km~s$^{-1}$ for parallel propagation and $U_2 \approx 12.64$~km~s$^{-1}$ for anti-parallel waves. Note that in our particular example the flow velocities needed for the coupling situation are realistic and within the range of typically observed velocities. However, if threads with very different physical properties and, therefore, with very different Alfv\'en and sound speeds are considered, the coupling flow velocities could be larger than the observed values. This means that the conditions necessary for collective oscillations of systems of threads with very different temperatures and/or densities may not be realistic in the context of solar prominences.

\section{Conclusion}
\label{sec:conclusions}

In this work, we have assessed the effect of mass flows on the collective behavior of slow and fast kink magnetosonic wave modes in systems of prominence threads. We have seen that the relation between the individual Alfv\'en (sound) speed of threads is the relevant parameter which determines whether the behavior of kink (slow) modes is collective or individual. In the absence of flows and when the Alfv\'en speeds of threads are similar, kink modes are of collective type. On the contrary, perturbations are confined within an individual thread if the Alfv\'en speeds differ. In the case of slow modes, the conclusion is equivalent but replacing the Alfv\'en speeds by the sound speeds of threads. On the other hand, when flows are present in the equilibrium, one can find again collective motions even in systems of non-identical threads by considering appropriate flow velocities. These velocities are within the observed values if threads with not too different temperatures and densities are assumed. However, since the flow velocities required for collective oscillations must take very particular values, such a special situation may rarely occur in real prominences. 

Therefore, if coherent oscillations of groups of threads are observed in prominences  \citep[e.g.,][]{lin2007}, we conclude that either the physical properties and flow velocities of all oscillating threads are quite similar or, if they have different properties, the flow velocities within threads are the appropriate ones to allow collective motions. From our point of view, the first option is the most probable one since the flow velocities required in the second case correspond to a very peculiar situation. This conclusion has important repercussions for future prominence seismological applications, in the sense that if collective oscillations are observed in large areas of a prominence, threads in such regions should possess very similar temperatures, densities, and magnetic field strengths

Here, we have only considered two-thread systems, but the method can be applied to an arbitrary multi-thread configuration. So, the model developed here could be used to perform seismological studies of large ensembles of prominence threads if future observations provide with positions and physical parameters of such systems.

\acknowledgements{
     R.~Soler is very grateful to M.~Luna for his useful comments and fruitful discussions during the preparation of the paper. The authors acknowledge the financial support received from the Spanish MCyT and the Conselleria d'Economia, Hisenda i Innovaci\'o of the CAIB under Grants No. AYA2006-07637 and PCTIB-2005GC3-03, respectively. R.~Soler also thanks the Conselleria d'Economia, Hisenda i Innovaci\'o for a fellowship. Finally, the authors want to acknowledge the International Space Science Institute teams ``Coronal waves and Oscillations'' and ``Spectroscopy and Imaging of quiescent and eruptive solar prominences from space'' for useful discussions.}

\end{document}